\documentclass[12pt]{article}
\usepackage{epsfig,amssymb,psfrag}

\catcode`\@=11
\def \thesection {\arabic{section}.}

\def \be  {\begin{equation}}
\def \ee  {\end{equation}}
\def \ba  {\begin{eqnarray}}
\def \ea  {\end{eqnarray}}
\def \baa {\begin{eqnarray*}}
\def \eaa {\end{eqnarray*}}
\def \bb  {\begin {thebibliography} }
\def \eb  {\end{thebibliography}}
\def \lab #1 {\label{#1}}
\newcommand \ci [1] {\cite{#1}}
\newcommand \bi [1] {\bibitem{#1}}
\newcommand\re[1]{(\ref{#1})}

\def \matrix #1 {\left(\begin{array}{cc} #1 \end{array}\right)}

\def \tr {\mathop{\rm tr}\nolimits}
\def \Im {\mathop{\rm Im}\nolimits}
\def \Re {\mathop{\rm Re}\nolimits}
\def \res{\mathop{\rm res}\nolimits}
\def \e  {\mathop{\rm e}\nolimits}
\newcommand\lr[1]{{\left({#1}\right)}}
\newcommand \widebar [1] {\overline{#1}}

\newcommand{\as}{\ifmmode\alpha_{\rm s}\else{$\alpha_{\rm s}$}\fi}
\newcommand{\asbar}{\ifmmode\bar{\alpha}_{\rm s}\else{$\bar{\alpha}_{\rm s}$}\fi}

\font\cmss=cmss12 
\def\inbar{\,\vrule height1.5ex width.4pt depth0pt}
\def\IC{\relax\hbox{$\inbar\kern-.3em{\rm C}$}}
\def\IZ{\relax{\hbox{\cmss Z\kern-.4em Z}}}
\def\IR{{\hbox{{\rm I}\kern-.2em\hbox{\rm R}}}}

\def\IP{{\hbox{{\rm I}\kern-.2em\hbox{\rm P}}}}
\def\II{\hbox{{1}\kern-.25em\hbox{l}}}

\def\numberbysection{\@addtoreset{equation}{section}
                     \def\theequation{\thesection\arabic{equation}}}
\numberbysection

\newcommand \Mybf[1] {\mbox{\boldmath$ {#1} $}}

\psfrag{a1}[cc][cc]{$\alpha_1$}
\psfrag{a2}[cc][bc]{$\alpha_{N-2}$}
\psfrag{b1}[cc][bc]{$\beta_1$}
\psfrag{b2}[cc][cc]{$\beta_{N-2}$}
\psfrag{g}[tc][br]{$\gamma_{P_0^-P_0^+}$}
\psfrag{d1}[cc][cc]{$\sigma_1$}
\psfrag{d2}[cc][cc]{$\sigma_2$}
\psfrag{d3}[cc][cc]{$\sigma_3$}
\psfrag{d4}[cc][cc]{$\sigma_{2N-4}$}
\psfrag{d5}[cc][cc]{$\sigma_{2N-3}$}
\psfrag{d6}[cc][cc]{$\sigma_{2N-2}$}
\psfrag{zero}[cc][cc]{$0$}

\psfrag{e1}[cc][cc]{$\Mybf{e}_1$}
\psfrag{e2}[cc][cc]{$\Mybf{e}_2$}
\psfrag{eN}[cc][cc]{$\Mybf{e}_N$}
\psfrag{qN}[cc][cc]{$q_N^{1/N}$}
\psfrag{dots}[cc][cc]{$\Huge\mathbf{\cdots}$}

\begin{document}

\begin{titlepage}
\begin{flushright}
\begin{tabular}{l}
LPT--Orsay--02--111\\
RUB--TP2--17/02\\
hep-th/0212169
\end{tabular}
\end{flushright}

\vskip3cm
\begin{center}
  {\large \bf
  Noncompact Heisenberg spin magnets from high-energy QCD  \\[2mm] III.~Quasiclassical approach}

\def\thefootnote{\fnsymbol{footnote}}%
\vspace{1cm}
{\sc S.\'{E}. Derkachov}${}^1$, {\sc G.P.~Korchemsky}${}^2$ and {\sc
A.N.~Manashov}${}^3$\footnote{ Permanent address:\ Department of Theoretical
Physics,  Sankt-Petersburg State University, 199034 St.-Petersburg, Russia}
\\[0.5cm]

\vspace*{0.1cm} ${}^1$ {\it
Department of Mathematics, St.-Petersburg Technology Institute,\\
198013 St.-Petersburg, Russia
                       } \\[0.2cm]
\vspace*{0.1cm} ${}^2$ {\it
Laboratoire de Physique Th\'eorique%
\footnote{Unite Mixte de Recherche du CNRS (UMR 8627)},
Universit\'e de Paris XI, \\
91405 Orsay C\'edex, France
                       }  \\[0.2cm]
\vspace*{0.1cm} ${}^3$
 {\it Institut f\"ur Theoretische Physik II, Ruhr--Universit\"at Bochum, \\
 44780 Bochum, Germany}

\vskip2cm
{\bf Abstract:\\[10pt]} \parbox[t]{\textwidth}{
The exact solution of the noncompact $SL(2,\mathbb{C})$ Heisenberg spin magnet
reveals a hidden symmetry of the energy spectrum. To understand its origin, we
solve the spectral problem for the model within quasiclassical approach. In this
approach, the integrals of motion satisfy the Bohr-Sommerfeld quantization
conditions imposed on the orbits of classical motion. In the representation of
the separated coordinates, the latter wrap around a Riemann surface defined by
the spectral curve of the model. A novel feature of the obtained quantization
conditions is that they involve both the $\alpha-$ and $\beta-$periods of the
action differential on the Riemann surface, thus allowing us to find their
solutions by exploring the full modular group of the spectral curve. We
demonstrate that the quasiclassical energy spectrum is in a good agreement with
the exact results.}
\vskip1cm

\end{center}

\end{titlepage}

\newpage

\tableofcontents

\newpage

\setcounter{footnote} 0

\section{Introduction}

Exact solution of the spectral problem for quantum-mechanical multi-particle
systems is the central problem in the theory of Integrable Lattice
Models~\cite{Bax}. One of the best known examples of such systems is spin$-1/2$
Heisenberg spin magnet. The model can be solved exactly by the Algebraic Bethe
Ansatz (ABA) and it has numerous applications~\cite{QISM,XXX,ABA}. The Heisenberg
magnet model can be generalized from the compact $SU(2)$ spins to noncompact
spins, ``living'' in an infinite-dimensional, unitary representations of the
$SL(2,\mathbb{C})$ group. The corresponding integrable model describes a
nearest-neighbour interaction between the $SL(2,\mathbb{C})$ spins and is called
the noncompact Heisenberg magnet~\cite{DKM-I}.

The noncompact Heisenberg spin magnets have important implications in high-energy
QCD~\cite{L1,FK}. It is well-known that hadronic scattering amplitudes grow as a
power of energy in agreement with the Regge model. In perturbative QCD framework,
this behaviour can be attributed to a contribution of colour-singlet gluonic
compound states. These states satisfy the Bartels-Kwiecinski-Praszalowicz
equation which coincides, in the multi-colour limit, with the Schr\"odinger
equation for the noncompact $SL(2,\mathbb{C})$ spin magnet. The effective QCD
interaction between $N$ reggeized gluons $(N=2,3,\ldots)$ occurs on
two-dimensional plane of transverse degrees of freedom (the impact parameter
space). It is described by the Hamiltonian $\mathcal{H}_N$, which defines a
quantum-mechanical system of $N$ particles with the coordinates $\vec
z_k=(x_k,y_k)$ and the momenta $\vec p_k=-i\vec\partial_{k}$, such that
$[z_k^\alpha,p_n^\beta]=i\delta_{kn}\delta^{\alpha\beta}$  ($k,n=1,\ldots,N$ and
$\alpha,\beta=1,2$) with the Planck constant $\hbar=1$. To map this system into a
Heisenberg magnet, one introduces holomorphic, $z=x+iy$, and antiholomorphic,
$\bar z=x-iy$, complex coordinates on the plane and defines the spin operators as
\be
S_k^0= iz_k p_k+s\,,\qquad S_k^-= -ip_k\,,\qquad S_k^+=iz_k^2p_k+2sz_k\,.
\label{spins}
\ee
Here $p_k=-i\partial/\partial z_k$ is a (complex) momentum along the
$z-$direction and the parameter $s=(1+n_s)/2+i\nu_s$ (with $n_s$ integer and
$\nu_s$ real) is a single-particle $SL(2,\mathbb{C})$ spin. One also defines the
antiholomorphic spin operators ${\bar S}_k^0$, ${\bar S}_k^-$ and ${\bar S}_k^+$
acting along the $\bar z-$direction. They are given by similar expressions with
$z_k$ and $s$ replaced by $\bar z_k=z_k^*$ and $\bar s=1-s^*=(1-n_s)/2+i\nu_s$,
respectively. The operators $\vec{S}_k$ and $\vec{\bar S}_n$ act along different
directions on the $\vec z-$plane and therefore commute. They satisfy the standard
$sl(2)$ commutation relations
\be
 [S_k^+,S_n^-]=2S_k^0 \delta_{kn}\,,\qquad
 [S_k^0,S_n^\pm]=\pm S_k^\pm\delta_{kn}\,,
\ee
with the quadratic Casimir operator $\vec
S_n^2=\lr{S_n^0}^2+\lr{S_n^+S_n^-+S_n^-S_n^+}/2=s(s-1)$. Similar relations hold
in the antiholomorphic sector. Defined in this way, the spin operators  $\vec
S_k$ and $\vec{\bar S_k}$ are the generators of the unitary principal series
representation of the $SL(2,\mathbb{C})$ group labelled by the pair of spins
$(s,\bar s)$, or equivalently integer $n_s$ and real $\nu_s$~\cite{group}. The
interaction between $N$ reggeized gluons in multi-colour QCD is translated into
the nearest-neighbour interaction between the spins $\vec S_k$ and $\vec{\bar
S_k}$ for the $SL(2,\mathbb{C})$ Heisenberg magnet at $s=0$ and $\bar
s=1$~\cite{L1,FK}
\be
{\cal H}_N=\sum_{k=1}^N H_{k,k+1}\,,\qquad H_{k,k+1}=H(\vec S_k\cdot
\vec S_{k+1})+ H(\vec{\bar S}_k\cdot\vec{\bar S}_{k+1})\,,
\label{H_N}
\ee
where $H_{N,N+1}=H_{N,1}$ and the two-particle Hamiltonian is expressed in terms
of the Euler $\psi-$function, $H(x)=\psi(j(x))+\psi(1-j(x))-2\psi(1)$ with
$j(j-1)=2x+2s(s-1)$. As follows from \re{H_N}, ${\cal H}_N$ can be split into a
sum of two mutually commuting Hamiltonians acting along the $z-$ and $\bar
z-$directions.

The noncompact $SL(2,\mathbb{C})$ Heisenberg magnet \re{H_N} is a completely
integrable model. It possesses a large enough set of mutually commuting conserved
charges $q_n$ and $\bar q_n$ $(n=2,...,N)$ such that $\bar q_n= q_n^\dagger$ and
$[\mathcal{H}_N,q_n]=[\mathcal{H}_N,\bar q_n]=0$. The charges $q_n$ are
polynomials of degree $n$ in the holomorphic spin operators. They have a
particular simple form at $s=0$~\cite{L1,FK}
\be
q_n= \sum_{1\le j_1 < j_2 < ... < j_n \le N} z_{j_1j_2}z_{j_2j_3}... z_{j_nj_1}
p_{j_1}p_{j_2}...p_{j_n}
\label{q_n}
\ee
with $z_{jk}=z_j-z_k$ and $p_j$ defined in \re{spins}. The ``lowest'' charge
$q_2$ is related to the total spin of the system $h$. For the principal series of
the $SL(2,\mathbb{C})$ it takes the following values~\cite{group}
\be
q_2=-h(h-1)+Ns(s-1)\,,\qquad h=\frac{1+n_h}2+i\nu_h\,,
\label{h}
\ee
with $n_h$ integer and $\nu_h$ real. The eigenvalues of the integrals of motion,
$q_2,...,q_N$, form the total set of quantum numbers parameterizing the
eigenstates of the model \re{H_N}. As was already mentioned, at $s=0$ and $\bar
s=1$ the latter define the multi-gluonic compound states in multi-colour QCD.

In spite of the fact that the noncompact $SL(2,\mathbb{C})$ Heisenberg magnet
represents a generalization of the compact $SU(2)$ spin chain, a very little has
been known about its energy spectrum till recently. One of the reasons is that
the exact solution of the eigenproblem for the Hamiltonian \re{H_N} represents a
difficulty of principle. In distinction with the compact magnets, the quantum
space of the $SL(2,\mathbb{C})$ magnet does not possess the highest weight and,
as a consequence, the conventional methods like the ABA method~\cite{QISM,ABA}
are not applicable.

The eigenproblem for the noncompact $SL(2,\mathbb{C})$ Heisenberg magnet has been
solved exactly in Refs.~\cite{DKM-I,KKM,DKKM} using the method the Baxter
$\mathbb{Q}-$operator~\cite{Bax}. This method allowed us to establish the
quantization conditions for the integrals of motion of the model, $q_3,...,q_N$,
obtain an explicit form of the dependence of the energy on the integrals of
motion, $E_N=E_N(\nu_h,n_h,q_3,...,q_N)$, and construct the corresponding
eigenfunctions in the representation of the Separated Variables~\cite{SoV}.
Solving the quantization conditions, we calculated the spectrum of the noncompact
$SL(2,\mathbb{C})$ magnet of spin $s=0$ for the number of particles $2\le N\le
8$. Its close examination revealed the following properties of the
model~\cite{DKKM}:
\begin{itemize}
\item Quantized values of the charges $q_k$ (with $k=3,...,N$)
depend on the ``hidden'' set of integers
$\Mybf{\ell}=(\ell_1,\ell_2,...,\ell_{2(N-2)})$
\be
q_k=q_k(\nu_h;n_h,\Mybf{\ell})\,,
\label{q-ell}
\ee
where integer $n_h$ and real $\nu_h$ define the total $SL(2,\mathbb{C})$ spin of
the state, Eq.~\re{h}.
\item As a function of continuous $\nu_h$, the charges
form the family of trajectories in the moduli space $\Mybf{q}=(q_2,q_3,...,q_N)$
labelled by integers $n_h$ and $\Mybf{\ell}$ . Each trajectory in the $q-$space
induces the corresponding trajectory for the energy $E_N$ (see
Figure~\ref{Fig:energy} below)
\be
E_N=E_N(\nu_h;n_h,\Mybf{\ell})\,.
\ee
\item For fixed total $SL(2,\mathbb{C})$ spin of the model, Eq.~\re{h},
the eigenvalues of the ``highest'' charge $q_N^{1/N}$ define an infinite set of
distinct points on the complex $q_N^{1/N}-$plane. At $N=3$ and $N=4$ they are
located close to the vertices of a lattice built from equilateral triangles and
squares, respectively (see Figures~\ref{Fig:q3} and \ref{Fig:q4} below).
\end{itemize}
Their origin remains obscure mainly due to a rather complicated form of the exact
quantization conditions. The main goal of this paper is to present a physical
interpretation of these properties. Our analysis is based on a generalization of
the well-known quasiclassical methods to noncompact Heisenberg magnets. One might
expect {\it a priori\/} that these methods could be applicable only for high
excited states. Nevertheless, as we will demonstrate below, the quasiclassical
formulae work with a good accuracy throughout the whole spectrum of the
noncompact $SL(2,\mathbb{C})$ Heisenberg magnet.

To formulate the quasiclassical solution of the eigenproblem for Hamiltonian
\re{H_N}, one has to consider a classical analog of the noncompact Heisenberg
spin magnet~\cite{Sol,GKK}. From point of view of classical dynamics, the model
describes a chain of $N$ interacting particles on the two-dimensional $\vec
z-$plane. We use (anti)holomorphic variables on the phase space and define the
coordinates and the momenta of particles as $\vec z_k=(z_k,\bar z_k)$ and $\vec
p_k=(p_k,\bar p_k)$, respectively. By the definition, $z_k$ and $p_k$ take
complex values such
that $\bar z_k=z_k^*$ and $\bar p_k=p_k^*$.%
\footnote{Of course, one can work instead with real, Cartesian coordinates,
but our choice is advantageous as it is dictated by the chiral structure of the
Hamiltonian \re{H_N}.} The only non-trivial Poisson bracket is given by
$\{z_k,p_n\}=\{\bar z_k,\bar p_n\}=\delta_{kn}$. The classical model inherits a
complete integrability of the quantum noncompact spin magnet. Its Hamiltonian and
the integrals of motion are obtained from \re{H_N}, \re{spins} and \re{q_n} by
replacing the momentum operators by the corresponding classical functions leading
to $\{q_k,{\cal H}_N\}=\{q_k,q_n\}=0$. Since the Hamiltonian \re{H_N} is given by
the sum of holomorphic and antiholomorphic functions, from point of view of
classical dynamics the model describes two copies of one-dimensional systems
``living'' on the complex $z-$ and $\bar z-$lines.
The solutions to the classical equations of motion have a rich structure and turn
out to be intrinsically related to the finite-gap solutions to the soliton
equations \cite{NMPZ,Kr}. Namely, the classical trajectories have the form of
soliton waves propagating in the chain of $N$ particles. Their explicit form in
terms of the Riemann $\theta-$functions was established in \cite{Sol} by the
methods of the finite-gap theory \cite{NMPZ,Kr}. The charges $\Mybf{q}$ define
the moduli of the soliton solutions and take arbitrary complex values in the
classical model. Going over to the quantum model, one finds that $\Mybf{q}$ are
quantized. In the quasiclassical approach presented in this paper, their values
satisfy the Bohr--Sommerfeld quantization conditions imposed on the orbits of
classical motion of $N$ particles.

In a standard manner, the WKB ansatz for the eigenfunction of the model \re{H_N}
involves the ``action'' function, $\Psi_{\rm WKB}(\vec z_1,\ldots,\vec z_N)\sim
\exp(iS_0/\hbar)$. 
Due to complete integrability of the classical system, it can be defined as a
simultaneous solution to the system of the Hamilton-Jacobi equations
\be
\sum_{k=1}^N \frac{\partial S_0}{\partial z_k}= P\,,\qquad
\mathrm{q}_n\lr{\Mybf{z},\frac{\partial S_0}{\partial \Mybf{z}}}= q_n\,,
\qquad
(n=2,...,N)\,,
\label{H-J}
\ee
where $\Mybf{z}=(z_1,...,z_N)$ denotes the set of holomorphic coordinates,
$\mathrm{q}_n(\Mybf{z},\Mybf{p})$ stands for the symbol of the operator \re{q_n}
and $P$ is a holomorphic component of the total momentum of $N$ particles. The
$\bar z-$dependence of $S_0$ is constrained by similar relations in the
antiholomorphic sector. To find a general solution to Eq.~\re{H-J}, one performs
a canonical transformation to the classical separated coordinates~\cite{SoV,NMPZ}
\be
(\vec z_1,\vec z_2,...,\vec z_{N})\ \stackrel{{\rm SoV}}{\mapsto} \ (\vec
z_0,\vec x_1,\vec x_2,...,\vec x_{N-1})\,,
\label{SoV-class}
\ee
with $\vec z_0$ the center-of-mass coordinate of the system and $\vec
x_n=(x_n,\bar x_n=x_n^*)$ new collective (separated) coordinates. As explained in
Section~2.1, the classical dynamics in the separated variables is determined by
the spectral curve (``equal energy'' condition)
\be
\Gamma_N: \qquad y^2=t_N^2(x) - 4x^{2N}\,,\qquad
t_N(x)=2x^N + q_2 x^{N-2} + ...+ q_{N-1} x + q_N\,,
\label{curve}
\ee
with $y(x)=2x^N\sinh p_x$ and $p_x$ being the momentum in the separated
coordinates. Here $t_N(x)$ is a polynomial of degree $N$ with the coefficients
defined by the holomorphic integrals of motion $q_n$. The spectral curve
establishes the relation between the holomorphic components of the separated
coordinates, $x$ and $p_x$, for a given set of the ``energies'' $q_2$, $\ldots$,
$q_N$. As we will see below, its properties play the central r\^ole in our
analysis.

In the separated coordinates, the solution to the Hamilton-Jacobi equations
\re{H-J} takes the form $S_0(\vec z_0,\vec x_1,\vec x_2,...,\vec x_{N-1})=(\vec
P\cdot
\vec z_0)+ \sum_{k=1}^{N-1}S_0(\vec x_k)$ with~\cite{NMPZ}
\be
S_0(\vec x) =\int^x_{x_0} dx\, p_x + \int^{\bar x}_{\bar x_0} d\bar x\, \bar
p_{\bar x} = 2\Re\int^x_{x_0} dx \, p_x\,.
\ee
Here complex momentum $\bar p_{\bar x}=p_x^*$ was defined in \re{curve} and $\vec
x_0$ is arbitrary. The WKB expression for the wave function in the separated
coordinates, $\sim\exp\lr{iS_0(\vec z_0,\vec x_1,\vec x_2,...,\vec
x_{N-1})/\hbar}$, factorizes into a product of single-particle wave functions,
$Q_{_{\rm WKB}}(\vec x_k)\sim
\exp\lr{{i} S_0(\vec x_k)/{\hbar}}$. According to \re{curve}, the momentum,
$p_x$, and, as a consequence, the action function $S_0(\vec x)$ are multi-valued
functions of $x$. Denoting the different branches of the action function as
$S_{0,\,\alpha}(\vec x)$, one writes the WKB expression for the wave function of
the quantum spin magnet as a sum over branches~\cite{PG,K97}
\be
Q_{\rm WKB}(\vec x) = \sum_\alpha A_\alpha(\vec x)
\exp\lr{\frac{i}{\hbar} S_{0,\,\alpha}(\vec x)}\,.
\label{WKB-wave}
\ee
The function $A_k(\vec x)$ takes into account subleading WKB corrections and is
uniquely fixed by $S_{0,\alpha}(\vec x)$. In general, the expression in the
r.h.s.\ of \re{WKB-wave} is not a single-valued function of $\vec x$. For $Q_{\rm
WKB}(\vec x)$ to be well-defined, the charges $\Mybf{q}$ have to satisfy the
Bohr-Sommerfeld quantization conditions. One of the main results in this paper is
that these conditions can be expressed in terms of the periods of the ``action''
differential over the canonical set of the $\alpha-$ and $\beta-$cycles on the
Riemann surface corresponding to the complex curve \re{curve}
\be
\Re \oint_{\alpha_k}dx\,p_x =\pi \hbar\,\ell_{2k-1}\,,\qquad
\Re \oint_{\beta_k}dx\,p_x =\pi \hbar\, \ell_{2k}\,,\qquad (k=1,..,N-2)\,,
\label{WKB-intro}
\ee
with $\Mybf{\ell}=(\ell_1,\ldots,\ell_{2N-4})$ being the set of integers.

The relations \re{WKB-intro} define the system of $2(N-2)$ real equations on the
$(N-2)$ complex charges $q_3,...,q_N$ (we recall that the eigenvalues of the
``lowest'' charge $q_2$ are given by \re{h}). Their solutions lead to the
quasiclassical expressions for the eigenvalues of the integrals of motion of the
noncompact spin magnet. As we will demonstrate in Section~5, these expressions
have the form \re{q-ell} and are in a good agreement with the exact results of
Ref.~\cite{KKM,DKKM}. A novel feature of the quantization conditions
\re{WKB-intro} is that they involve {\it both\/} the $\alpha-$ and
$\beta-$periods on the Riemann surface. This should be compared with the
situation in one-dimensional lattice integrable models, like the Toda chain
model~\cite{PG,FS} and the $SL(2,\mathbb{R})$ Heisenberg spin
magnet~\cite{K2,BDKM,AB}. There, the WKB quantization conditions involve only the
$\alpha-$cycles, since the $\beta-$cycles correspond to classically forbidden
zones. In the $SL(2,\mathbb{C})$ spin magnet, the classical trajectories wrap
over an arbitrary closed contour on the spectral curve \re{curve} leading to
\re{WKB-intro}. This fact allows one to explore the full modular group~\ci{D} of
the complex curve \re{curve}.

The paper is organized as follows. In Section~2 we remind the definition of the
noncompact Heisenberg spin magnet both in the classical and quantum cases. Going
over to the representation of the Separated Variables we construct the wave
function of the model in terms of the solutions to the Baxter equation. Applying
the WKB methods, we solve the Baxter equation in Section~3 and show that the
quasiclassical expression for the wave function is uniquely defined by the
complex curve $\Gamma_N$ introduced in \re{curve}. Requiring this function to be
single-valued, we obtain the quantization conditions \re{WKB-intro} for the
integrals of motion $q_n$. In Section 4 we obtain quasiclassical expressions for
the energy and the quasimomentum. Both observables are expressed in terms of the
$Q-$blocks, which satisfy the holomorphic Baxter equation and have prescribed
analytical properties and asymptotic behaviour at infinity. In Section 5 we
analyze the quantization conditions \re{WKB-intro} and compare their solutions
with the exact results for the energy spectrum. Section 6 contains concluding
remarks. Some technical details of the calculations are summarized in two
Appendices.

\section{Noncompact Heisenberg spin magnet}

Let us summarize, following \cite{Sol,GKK}, the main features of the
$SL(2,\mathbb{C})$ Heisenberg spin magnet in the classical and quantum mechanics.

\subsection{Classical model}

In the classical case, the model describes the chain of $N$ interacting particles
on the two-dimensional plane with the Hamiltonian \re{H_N}. The classical motion
along the complex $z-$direction is described by the Hamilton equations
\be
\partial_t z_k = \{ z_k, \mathcal{H}_N\}=\frac{\partial \mathcal{H}_N}{\partial p_k}
\,,\qquad
\partial_t p_k = \{ p_k, \mathcal{H}_N\}=-\frac{\partial \mathcal{H}_N}{\partial
z_k}\,.
\label{Ham-eq}
\ee
This system is completely integrable and the integrals of the motion are given by
\re{q_n}. Following the Quantum Inverse Scattering Method~\cite{QISM}, one can
describe the classical Heisenberg spin magnet by the Lax matrix
\be
L_k(u)=u\cdot \II+ iS_k^0\cdot \sigma^3 + iS_k^+\cdot \sigma^- + iS_k^-\cdot
\sigma^+=\lr{\begin{array}{cc}
  u+iS_k^0 & iS_k^- \\
  iS_k^+ & u-iS_k^0
\end{array}  }\,,
\ee
with $\sigma^a$ being the Pauli matrices. The dynamical variables $S_k^0$,
$S_k^-$ and $S_k^+$ depend on the (holomorphic) coordinates and momenta of
particles, $z_k$ and $p_k$, respectively, and they are given by the same
expressions as in \re{spins}. The Hamilton equations \re{Ham-eq} are equivalent
to the matrix Lax pair relation
\be
\partial_t L_k(u) = \{ L_k(u),\mathcal{H}_N\}= A_{k+1}(u) L_k(u) - L_k(u)
A_k(u)\,,
\ee
with $A_k(u)$ being a $2\times 2$ matrix depending on the coordinates and momenta
of particles.

The exact integration of the classical equations of motion is based on the
Baker-Akhiezer function $\Psi_k(u;t)$~\cite{Kr}. By the definition, it satisfies
the system of matrix relations
\be
L_k(u)\Psi_k(u;t)=\Psi_{k+1}(u;t)\,,\qquad \partial_t \Psi_k(u;t)= A_k(u)
\Psi_k(u;t)\,.
\ee
Introducing the monodromy matrix as a consecutive product of the Lax matrices,
one finds that it produces the shift of the Baker-Akhiezer function along the
chain
\be
T_N(u)=L_N(u)\ldots L_1(u)\,,\qquad T_N(u) \Psi_1(u;t)=\Psi_{N+1}(u;t)\,.
\label{T_N}
\ee
For periodic boundary conditions, $z_{k+N}=z_k$ and $p_{k+N}=p_k$, the
Baker-Akhiezer function satisfies the Bloch-Floquet relation
\be
\Psi_{N+k}(u;t)=w(u) \Psi_k(u;t)\,,
\ee
According to \re{T_N}, the Bloch-Floquet factor $w(u)$ is an eigenvalue of the
monodromy matrix. Therefore, it does not depend on the time and satisfies the
characteristic equation
\be
\det\lr{T_N(u)-w}=w^2-w\,t_N(u) +u^{2N}=0\,.
\label{char-eq}
\ee
Here $t_N(u)=\tr T_N(u)$ is a polynomial in $u$ of degree $N$ with the
coefficients given by the integrals of motion, Eq.~\re{curve}. Introducing the
complex function $y(u)=w-u^{2N}/w$, one obtains from \re{char-eq} that $y(u)$
defines the algebraic complex curve \re{curve}.

The Baker-Akhiezer function $\Psi_k(u;t)$ is a double-valued function on the
complex $u-$plane~\cite{Kr}. Its explicit expression in terms of the
theta-functions defined on the curve \re{curve} can be found in \cite{Sol}. We do
not present it here since we will not use the Baker-Akhiezer function in the rest
of the paper. The function $\Psi_k(u;t)$ has $N-1$ simple poles at $u=x_k$
($k=1,\ldots,N-1$) and the same number of zeros. Remarkable property of its poles
is that the variables $(x_k,p(x_k))$ (with $p(u)=\ln (w(u)/u^N)$ and
$k=1,\ldots,N-1$) form the set of holomorphic separated variables for the
classical model, Eq.~\re{SoV-class}. Notice that $x_k$ and $p(x_k)$ take
arbitrary complex values. This allows one to integrate the equations of motion
exactly and reconstruct the classical trajectories of particles on the Riemann
surface defined by the curve \re{curve}. The same classical motion describes a
soliton wave propagating in the chain of $N$ particles on the two-dimensional
$\vec z-$plane.

\subsection{Quantum model}

In the quantum case, the eigenfunction of the $SL(2,\mathbb{C})$ Heisenberg spin
magnet, $\Psi(\vec z_1,\ldots,\vec z_N)$, is defined as a simultaneous eigenstate
of the integrals of motion $q_2,\ldots,q_N$, Eq.~\re{q_n}, and their
antiholomorphic counterparts. Together with the total momentum of the system,
$\vec P$, their eigenvalues $q=(q_2,\ldots,q_N)$ define the total set of the
quantum numbers of the model. Due to chiral structure of the Hamiltonian and the
integrals of motion, the eigenfunction can be decomposed as~\cite{JW}
\be
\Psi(\vec z_1,\ldots,\vec z_N)=\sum_{a,\,b}C_{ab}(q,\bar q) \Psi_q^{(a)}(z_1,\ldots,z_N)
\widebar \Psi_{\bar q}^{(b)}(\bar z_1,\ldots,\bar z_N)\,,
\label{mixing}
\ee
where $\Psi_q^{(a)}$ and $\widebar \Psi_{\bar q}^{(b)}$ diagonalize the integrals
of motion in the holomorphic and antiholomorphic sectors, respectively, and
$C_{ab}$ are mixing coefficients. The function $\Psi(\vec z_1,\ldots,\vec z_N)$
is a single-valued function on the two-dimensional plane. It belongs to the
principal series of the $SL(2,\mathbb{C})$ group labelled by the spins $(h,\bar
h)$ defined in \re{h} (with $\bar h=1-h^*$) and is normalizable with respect to
the $SL(2,\mathbb{C})$ scalar product. In contrast with $\Psi(\vec
z_1,\ldots,\vec z_N)$, the chiral solutions $\Psi_q^{(a)}$ and $\widebar
\Psi_{\bar q}^{(b)}$ acquire nontrivial monodromy when $\vec z_k$ encircles
other particles on the plane. The quantization conditions for the integrals of
motion $\Mybf{q}$ follow from the requirement that the monodromy should cancel in
the r.h.s.\ of \re{mixing}. The same condition fixes (up to an overall
normalization) the mixing coefficients $C_{ab}$.

To formulate the quantization conditions it is convenient to switch from the
coordinate $\vec z-$representation to the representation of the Separated
Variables (SoV)~\cite{SoV}. In this representation, the wave function takes a
factorized form
\be
\Psi(\vec z_1,\vec z_2,...,\vec z_{N}) \stackrel{{\rm SoV}}{\longrightarrow}
\Phi(\vec z_0,\vec x_1,\vec x_2,...,\vec x_{N-1})=\e^{i\vec P\cdot\vec
z_0}\,  Q(\vec x_1)\,Q(\vec x_2) \ldots Q(\vec x_N)\,,
\label{SoV}
\ee
where $\vec P$ is the total momenta of $N$ particles, $\vec z_0$ is the
center-of-mass coordinate of the system and $\vec x_1,\ldots,\vec x_{N-1}$ are
new collective (separated) coordinates.  The explicit form of the unitary
transformation to the SoV representation can be found in \cite{DKM-I} (see also
\cite{dVL-I} for similar expressions at $N=2$ and $N=3$). The eigenfunction in
the SoV representation has the following properties. Introducing holomorphic and
antiholomorphic components $\vec x=(x,\bar x)$, one finds that the possible
values of the separated coordinates can be parameterized by integer $n$ and real
$\nu$ as
\be
x=\nu-\frac{in}2\,,\qquad \bar x=\nu+\frac{in}2\,,
\label{x}
\ee
so that $\bar x=x^*$ and $i(x-\bar x)=n$. Here, as before, one has $\hbar=1$. To
restore the $\hbar-$dependence one has to substitute $n\to\hbar n$. Notice that,
in contrast with the classical case, the separated variables have a discrete
imaginary part for finite $\hbar$.

A single-particle  $Q-$function entering \re{SoV} satisfies the holomorphic
Baxter equation
\be
(x+is)^N Q(x+i,\bar x)+(x-is)^N Q(x-i,\bar x)=t_N(x)\, Q(x,\bar x)\,,
\label{Bax-eq}
\ee
with $t_N(x)$ defined in \re{curve}. Similar equation holds for $Q(x,\bar x)$ in
the antiholomorphic sector with $s$ and $q_n$ replaced by $\bar s=1-s^*$ and
$\bar q_n=q_n^*$, respectively. The solution to the Baxter equations,
$Q(\nu-{in}/2,\nu+{in}/2)$, is a well-defined, regular function of integer $n$
and real $\nu$. At large $\nu$ and fixed $n$ it has the following asymptotic
behaviour
\be
Q(\nu-{in}/2,\nu+{in}/2) \sim \e^{i\Theta}\nu^{h+\bar h-N(s+\bar s)}+
\e^{-i\Theta}\nu^{1-h+1-\bar h-N(s+\bar s)}\,,
\label{as-pre}
\ee
where $h$ and $\bar h=1-h^*$ define the total $SL(2,\mathbb{C})$ spin of the
model, Eq.~\re{h}, and $\Theta$ is some phase.

The exact solution to Eqs.~\re{Bax-eq} and \re{as-pre} was constructed in
Refs.~\cite{KKM,DKKM}. It allowed us to establish the quantization conditions for
the integrals of motion and calculate the energy spectrum of the model. In this
paper we shall present another, quasiclassical approach to solving the Baxter
equations. Although it does not provide the exact solution for the $Q-$function,
it allows us to elucidate a hidden symmetry of the energy spectrum.

\section{Quasiclassical wave function}

The quasiclassical approach relies on the observation that the holomorphic Baxter
equation \re{Bax-eq} resembles a one-dimensional discrete Schr\"odinger equation.
A specific feature of the model is that $x-$coordinates entering the Baxter
equation takes complex values \re{x} and the
Planck constant equals unity $\hbar=1$.%
\footnote{In quantum models, like the Toda chain, the Planck constant controls a shift
of the argument of the $Q-$function in the l.h.s.\ of the Baxter equation
\re{Bax-eq}.} As a consequence, the quasiclassical limit corresponds to large
values of the ``energies'' $q_2,\ldots,q_N$ in Eq.~\re{curve}.

To perform the large $\Mybf{q}-$limit in \re{Bax-eq}, one introduces an arbitrary
auxiliary parameter $\eta$ and rescales simultaneously the coordinates, $x\to
x/\eta$, and the charges, $q_k \to q_k \eta^k$. Defining
\be
\widehat t_N(x)=\eta^N t_N(x/\eta)= 2x^N+\widehat q_2 x^{N-2} + ... + \widehat
q_N\,,
\label{t-hat}
\ee
with $\widehat q_n\equiv q_n \eta^n={\cal O}(\eta^0)$ as $\eta\to 0$, one finds
that this transformation allows one to get rid of large parameters in the Baxter
equation \re{Bax-eq}. At the same time, it sets the Planck constant as
$\hbar=\eta$. Let us look for the solution to the holomorphic Baxter equation in
the WKB form~\cite{PG,K97}
\be
Q(x/\eta)=\exp\lr{\frac{i}{\eta}\int_{x_0}^x dx\, S'(x)}\,,\qquad
S(x)=S_0(x)+\eta S_1(x) + {\cal O}(\eta^2)\,,
\label{WKB-exp}
\ee
where $S'(x)=d S(x)/dx$ and $x_0$ is an arbitrary reference point. Its
substitution into \re{Bax-eq} leads to the following relations
\be
2\cosh S_0'(x)=\frac{\widehat t_N(x)}{x^N}\,,\qquad S_1'(x)=\frac{i}2
\lr{\ln\sinh S_0'(x)}'+\frac{isN}{x}\,.
\label{S's}
\ee
One can systematically improve the WKB expansion \re{WKB-exp} and express
subleading corrections to $S(x)$ in terms of the leading term $S_0(x)$. Similar
expressions can be obtained for solutions to the antiholomorphic Baxter equation
$\widebar Q(\bar x)$. Then, a general WKB expression for $Q(x,\bar x)$ is given
by a bilinear combination of $Q(x)$ and $\widebar Q(\bar x)$.

\subsection{Properties of the WKB expansion}

Introducing notation for
\be
p_x=S_0'(x)\,,\qquad y(x)
=2x^N\sinh p_x\,,
\ee
one rewrites the first relation in \re{S's} as
\be
y^2(x)=\widehat{t}_N^{\ 2}(x)-4x^{2N}\,.
\label{w}
\ee
We notice that, up to rescaling of parameters, $y(x)$ coincides with the
hyperelliptic curve $\Gamma_N$, Eq.~\re{curve}. The coincidence is not accidental
of course. The leading term of the WKB expansion, $S_0(x)$, satisfies the
classical Hamilton-Jacobi equations in the separated coordinates. Its derivative,
$p_x=S_0'(x)$, defines a (complex-valued) holomorphic component of the momentum
in the separated variables. As such, it belongs to the spectral curve of the
classical model \re{char-eq} for $w(x)=x^N \exp(p_x)$.

The leading term of the WKB expansion \re{WKB-exp} can be calculated as
\be
S_0(x)=\int_{x_0}^x 
dx\,p_x= \int_{x_0}^x\frac{dx}{y(x)}
\left[N \widehat t_N(x)-x \widehat t_N'(x)\right]+xp_x\bigg|^{x}_{x_0}\,.
\label{dS_0}
\ee
Solving \re{w} we find that $y(x)$ and, as a consequence $S_0'(x)$, are
double-valued functions on the complex $x-$plane. To specify two branches of
$S_0'(x)$, one makes cuts on the $x-$plane in an arbitrary way between the
$2(N-1)$ branching points $\sigma_j$. The latter are defined as $y(\sigma_j)=0$,
or equivalently
\be
\widehat t_N^2(\sigma_j)-4\sigma_j^{2N}=(\widehat q_2 \sigma_j^{N-2}+...+\widehat q_N)
(4\sigma_j^N+\widehat q_2 \sigma_j^{N-2}+...+\widehat q_N)=0\,.
\label{br_points}
\ee
According to their definition, the branching points correspond to the special
points on the phase space of the classical system, in which the holomorphic
component of the momentum (in the separated coordinates) takes the values $p_x=0$
and $p_x=\pm i\pi$. Two different solutions to \re{w} give rise to two branches
$S_{0,+}(x)$ and $S_{0,-}(x)$ which are continuous functions of complex $x$
except across the cuts. These functions are transformed one into another as $x$
encircles the branching point $\sigma_j$ in the anticlockwise direction
\be
S_{0,\pm}'(x)\ \stackrel{x\circlearrowleft
\sigma_j}{\longrightarrow}\ -S_{0,\mp}'(x)\,.
\label{S-disc}
\ee
Their asymptotic behaviour at infinity can be found from \re{dS_0} and \re{w} as
\be
S_{0,\pm}'(x) \sim \pm \frac{{\widehat q_2}^{\,1/2}}{x} \sim \pm \frac{i}{x}
\eta(h-1/2)\,,
\label{S0-as}
\ee
as $x\to\infty$. Here, in the last relation, we replaced $\widehat q_2=q_2\eta^2$
by its expression, Eq.~\re{h}, and took the limit $\eta\to 0$ with
$|\eta(h-1/2)|=|\eta(i\nu_h+n_h/2)|$=fixed. Notice that the integration contour
in \re{dS_0} does not cross the cuts.

It becomes convenient to combine the two branches $S_{0,\pm}'(x)$ and define
$S_0'(x)$ as a single-valued function on the Riemann surface $\Gamma_N$ obtained
by gluing together two copies of the $x-$plane along the $(N-1)-$cuts
$[\sigma_{2j},\sigma_{2j+1}]$ running between the branching points. By the
definition, $S_0'(x)=S_{0,+}'(x)$ on one plane (upper sheet) and
$S_0'(x)=S_{0,-}'(x)$ on another one (lower sheet). Then, it follows from
Eq.~\re{dS_0} that $dx\,S_0'(x)$ is a well-defined, meromorphic differential on
$\Gamma_N$ of the third kind (the dipole differential). It has a pair of poles
located above the point $x=\infty$ on the upper and lower sheets, $P_\infty^+$
and $P_\infty^-$, respectively. Here we used the standard notation for the points
on the Riemann surface, $P_x^\pm=(x,\pm)$.

Substituting \re{dS_0} into \re{S's}, we calculate the first nonleading WKB
correction as
\be
S_1'(x)=\frac{i}2\lr{\ln \frac{y(x)}{2x^{N}}}'+\frac{iNs}{x} =
\frac{i}{4}\sum_{j=1}^{2N-2}\frac1{x-\sigma_j}+\frac{iN}{x}\lr{s-\frac12}\,.
\label{dS_1}
\ee
In distinction with the leading  $S_0-$term, the function $S_1'(x)$ is
well-defined on the complex $x-$plane. Therefore, it takes the same value on the
both sheets of the Riemann surface $\Gamma_N$ and its asymptotic behaviour for
$x\to\infty$ is given by
\be
S_1'(x) \sim \frac{i}{x}\lr{Ns-\frac12}\,.
\label{S1-as}
\ee

Combining together \re{dS_0} and \re{dS_1} we find that the two different
branches $S_{0,\pm}'(x)$, or equivalently two different sheets of the Riemann
surface $\Gamma_N$, give rise to two independent WKB solutions to the holomorphic
Baxter equation \re{Bax-eq}
\be
Q_\pm(x/\eta)=\exp\lr{\frac{i}{\eta}\int_{x_0}^x dx\,S_\pm'(x)}\,,\qquad
S_\pm'=S_{0,\pm}'(x)+\eta\,S_1'(x)+{\cal O}(\eta^2)\,.
\label{Q-WKB}
\ee
Their asymptotics at infinity can be obtained from \re{S0-as} and \re{S1-as} as
\be
Q_+(x/\eta)\sim x^{1-h-Ns}\,,\qquad Q_-(x/\eta)\sim x^{h-Ns}\,,
\label{Q-as}
\ee
as $x\to\infty$. Going over through similar analysis of the antiholomorphic
Baxter equation, one arrives at the WKB expressions for $\widebar Q_\pm(\bar
x/\eta)$. They can be obtained from \re{Q-WKB} by replacing holomorphic variables
by their counterparts  in the antiholomorphic sector. In this way, one gets from
\re{Q-as}
\be
\widebar Q_+(\bar x/\eta)\sim \bar x^{1-\bar h-N\bar s}\,,\qquad
\widebar Q_-(\bar x/\eta)\sim \bar x^{\bar h-N\bar s}\,,
\label{Qbar-as}
\ee
where $\bar x=x^*$, $\bar s=1-s^*$ and $\bar h=1-h^*$.

\subsection{Quantization conditions}

Let us construct the quasiclassical solution to the Baxter equation \re{Bax-eq}
as a bilinear combination of the chiral solutions $Q_\pm(x/\eta)$ and $\widebar
Q_\pm(\bar x/\eta)$
\be
Q(x/\eta,\bar x/\eta) = c_+\,Q_+(x/\eta)\widebar Q_+(\bar x/\eta)
+c_-\,Q_-(x/\eta)\widebar Q_-(\bar x/\eta)\,.
\label{Q-gen}
\ee
Using \re{Q-as} and \re{Qbar-as} one verifies that the wave function defined in
this way has correct asymptotic behaviour at infinity, Eq.~\re{as-pre}. The
cross-terms $Q_\pm \widebar Q_\mp$ do not enter \re{Q-gen} since they do not
verify \re{as-pre}. The functions $Q_\pm(x/\eta)$ and $\widebar Q_\pm(\bar
x/\eta)$ depend on the reference point $x_0$, Eq.~\re{Q-WKB}, whereas
$Q(x/\eta,\bar x/\eta)$ should not depend on the choice of $x_0$. This fixes the
$x_0-$dependence of the coefficients $c_\pm(x_0)$ as ($\bar x_0=x_0^*$)
\be
c_\pm(x_0')=c_\pm(x_0)\exp\lr{\frac{i}{\eta}
\left[\int_{x_0}^{x_0'}dx\,S_\pm'(x)+\int_{\bar x_0}^{\bar x_0'}
d\bar x\,\bar S_\pm'(\bar x)\right]}\,.
\label{AA-rel}
\ee

By the construction, the WKB formula \re{Q-gen} is valid for small $\eta$ and
$x\sim \eta^0$. It describes the wave function, $Q(x,\bar x)$, for large values
of the separated coordinates, $x\sim\bar x \sim 1/\eta$, or equivalently $n\sim
\nu\sim 1/\eta$ in the parameterization \re{x}. In this region, one can ignore
the fact that $n$ takes strictly integer values and treat the separated
coordinates $x$ and $\bar x$ as continuous complex, mutually conjugated
variables, $\bar x=x^*$. We recall that in one-dimensional lattice models, the
Toda chain~\cite{PG,FS} and the $SL(2,\mathbb{R})$ magnet~\cite{K2,BDKM,AB}, the
classical motion in the separated coordinates is restricted to finite intervals
on the real $x-$axis. The WKB wave function is a continuous, single-valued
function of real $x$, oscillating inside these intervals and vanishing at
infinity. Going back to the noncompact $SL(2,\mathbb{C})$ magnet, one finds that,
in distinction with the models mentioned above, the classical motion in the
separated coordinates occurs on the whole two-dimensional $\vec x-$plane. This
suggests that the WKB wave function $Q(x,\bar x=x^*)$ has to be well-defined on
the complex $x-$plane. In particular, contrary to the chiral solutions to the
Baxter equation \re{S-disc}, it should have a trivial monodromy around the
branching points $\sigma_j$. In other words, $Q(x,x^*)$ has to be a single-valued
function on the complex $x-$plane rather than on the Riemann surface $\Gamma_N$.
The former condition is much stronger than the latter one and, as we will show
below, it leads to the WKB quantization conditions for the integrals of motion
$\Mybf{q}$.

Let us examine the monodromy of the chiral solutions to the Baxter equation,
$Q_\pm(x/\eta)$, around the branching points $\sigma_j$, Eq.~\re{br_points}.
Encircling the branching point $\sigma_j$ on the $x-$complex plane, one finds
that the leading WKB term $S_0'(x)$ is transformed according to \re{S-disc} while
the subleading term $S_1'(x)$ stays invariant. Let us explore a freedom in
choosing the reference point $x_0$ in \re{Q-WKB} and put $x_0=\sigma_j$ in order
to ensure that $S_\pm(\sigma_j)=0$. Then, it follows from \re{S-disc} and
\re{Q-WKB} that the WKB solutions $Q_\pm(x)$ defined in this way are transformed
one into another as $x$ encircles $\sigma_j$ on the complex plane
\be
Q_\pm(x/\eta) 
\ \stackrel{x\circlearrowleft
\sigma_j}{\longrightarrow}\
Q_\mp (x/\eta)\,,\qquad
\mbox{(for $x_0=\sigma_j$)}\,.
\ee
Similar relations hold for the antiholomorphic solutions $\widebar Q_\pm(\bar
x/\eta)$ at $\bar x_0=\bar\sigma_j\equiv\sigma_j^*$. We find from \re{Q-gen} that
$Q(x/\eta,\bar x/\eta)$ stays invariant under this transformation provided that
\be
c_+(\sigma_j)=c_-(\sigma_j)\,,\qquad (j=1,2,...,2(N-1))\,.
\label{AA}
\ee
These conditions ensure that the quasiclassical wave function \re{Q-gen} is a
single-valued function on the complex $x-$plane. We recall that the branching
points are defined as solutions to Eq.~\re{br_points}.

For different $x_0$, the coefficients $c_\pm(x_0)$ are related to each other
according to \re{AA-rel}. Therefore, choosing $x_0'=\sigma_j$ and $x_0=0$ we
obtain from Eqs.~\re{AA-rel} and \re{AA}
\be
\frac{c_+(0)}{c_-(0)}
=\exp\lr{\frac{2i}{\eta}\Re\int_{0}^{\sigma_j}dx
\left[S_{0,-}'(x)-S_{0,+}'(x)\right]+{\cal O}(\eta)}\,,
\label{A-ratio}
\ee
where $j=1,...,2(N-1)$ and the integration contour does not cross the cuts on the
complex $x-$plane.
In arriving at \re{A-ratio}, we applied \re{Q-WKB} and took into account that
$\bar S_{0,\pm}'(\bar x)=\lr{S_{0,\pm}'(x)}^*$. Notice that the exponent in the
r.h.s.\ of \re{A-ratio} does not receive the $\mathcal{O}(\eta^0)-$correction,
since $S_1'(x)$ is single-valued on the $x-$plane. Since the l.h.s.\ of
\re{A-ratio} does not depend on $j$, one gets from \re{A-ratio} the set of
consistency conditions
\be
\exp\lr{\frac{2i}{\eta}\Re \int_{\sigma_k}^{\sigma_j}dx\left[S_{0,-}'(x)
-S_{0,+}'(x)\right]+{\cal O}(\eta)}=1\,.
\label{A-cons}
\ee
The quantization conditions \re{A-ratio} and \re{A-cons} can be expressed in a
concise form in terms of the contour integrals on the Riemann surface $\Gamma_N$,
Eq.~\re{curve}. We recall that the two branches $S_{0,\pm}'(x)$ define the dipole
differential $dx\,S_0'(x)$ on $\Gamma_N$. This allows one to rewrite \re{A-ratio}
as
\be
\frac{c_+(0)}{c_-(0)}
=\exp\lr{-\frac{2i}{\eta}\Re \int_{P_0^-}^{P_0^+}dx\,S_0'(x)+{\cal
O}(\eta)}\equiv
\e^{-2i\Theta}\,,
\label{A-curve}
\ee
with $\Theta$ some (real) constant introduced for later convenience. Here
integration goes over an arbitrary path on $\Gamma_N$, which starts at the point
$P_0^-$ located above $x=0$ on the lower sheet and ends at the point $P_0^+$
above $x=0$ on the upper sheet.

\begin{figure}[th]
\vspace*{5mm}
\centerline{{\epsfysize6cm \epsfbox{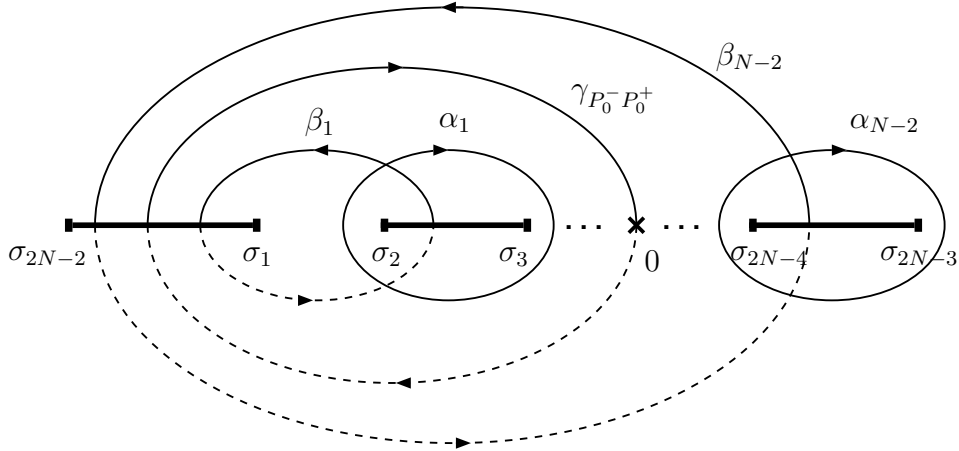}}}
\caption[]{The canonical basis of oriented $\alpha-$ and $\beta-$cycles on the Riemann
surface $\Gamma_N$. The dotted line represents the part of the $\beta-$cycles on
the lower sheet. The cross denotes a projection of the points $P_0^\pm$ onto the
complex plane. The path $\gamma_{P_0^-P_0^+}$ goes from the point $P_0^-$ on the
lower sheet to the point $P_0^+$ on the upper sheet.}
\label{Fig-cuts}
\end{figure}

Let us define the canonical basis of oriented cycles on $\Gamma_N$ as shown in
Figure~\ref{Fig-cuts}. The contour integral in Eq.~\re{A-curve} can be uniquely
decomposed over this basis as
\be
\int_{P_0^-}^{P_0^+}=\int_{\gamma_{P_0^-P_0^+}}+ \sum_{j=1}^{N-2} k_j
\oint_{\alpha_j} + \sum_{j=1}^{N-2} m_j\oint_{\beta_j}\,,
\label{path}
\ee
with $k_j$ and $m_j$ arbitrary integer. Here, the path $\gamma_{P_0^-P_0^+}$ goes
from the point $P_0^-$ on the lower sheet to the point $P_0^+$ on the upper sheet
and it does not cross the canonical cycles (see Figure~\ref{Fig-cuts}). The
l.h.s.\ of \re{A-curve} should not depend on the choice of the integration path,
or equivalently on integers $k_j$ and $m_j$ in Eq.~\re{path}. This leads to
\be
\Re\oint_{\alpha_k}dx\,S_0'(x)/\eta=\pi\ell_{2k-1}\,,\qquad
\Re\oint_{\beta_k}dx\,S_0'(x)/\eta=\pi\ell_{2k}\,,
\label{quan-cond}
\ee
with $S_0(x)$ defined in \re{dS_0}, $k=1,...,N-2$ and
$\Mybf{\ell}=(\ell_1,\ldots,\ell_{2N-4})$ being integer. These relations
establish the WKB quantization conditions for the integrals of motion of the
model. In addition, one finds from \re{path} and \re{A-curve}
\be
\Re \int_{\gamma_{P_0^-P_0^+}}dx\,S_0'(x)/\eta=\Theta\,.
\label{Theta}
\ee
The following comments are in order.

Eqs.~\re{quan-cond} and \re{Theta} are valid up to corrections suppressed by a
{\it second\/} power of $\eta$. One can show that this feature is rather general
and the WKB expansion in the l.h.s.\ of \re{quan-cond} and \re{Theta} goes over
{\it even\/} powers of $\eta$.

We recall that, by the definition, $\eta$ is a small auxiliary parameter which
was introduced in \re{t-hat} in order to formulate the WKB expansion. The
spectrum of the model should not depend on $\eta$.
Indeed, one verifies using \re{dS_0} and \re{t-hat}, that the l.h.s.\ of
\re{quan-cond} and \re{Theta}, as well as $q_n=\widehat q_n/\eta^n$, stay
invariant under the scaling transformation $x\to \lambda x$, $\eta \to
\lambda\eta$ and $\widehat q_n\to\widehat q_n
\lambda^n $.
Therefore we may choose $\eta$ in Eqs.~\re{quan-cond} and \re{Theta} to our best
convenience,  say $\eta=1$, but keep in mind, of course, that the WKB
approximation is valid for large values of the charges $q_n$. In this way, one
arrives at the quantization conditions \re{WKB-intro}. Their solutions have the
form \re{q_n} and are parameterized by the set of integers
$\Mybf{\ell}=(\ell_1,...,\ell_{2(N-2)})$.

As we will show in Section~4, the phase $\Theta$ in the r.h.s.\ of \re{Theta} is
closely related to the quasimomentum $\theta_N$ corresponding to the eigenstate
\re{Q-gen}. Evaluating the contour integral in the l.h.s.\ of \re{Theta} and
replacing the charges $q_n$ by their quantized values, Eq.~\re{q_n}, one can
obtain from \re{Theta} the dependence of the quasimomentum on integers
$\Mybf{\ell}$.

A novel feature of the quantization conditions \re{quan-cond} compared to the
conventional WKB approach \cite{PG,FS,Ahn} is that they involve {\it both\/}
$\alpha-$ and $\beta-$periods on the Riemann surface $\Gamma_N$. We remind that
the Hamiltonian of the $SL(2,\mathbb{C})$ magnet is given by the sum of two
one-dimensional mutually commuting Hamiltonians ``living'' on the $z-$ and $\bar
z-$lines. If we $z$ and $\bar z$ have been {\it real\/} coordinates, each of
these Hamiltonian would have defined quantum $SL(2,\mathbb{R})$ Heisenberg
magnet. The WKB quantization conditions for this magnet look as
follows~\cite{K97,BDKM,AB}
\be
SL(2,\mathbb{R}):\qquad \oint_{\alpha_k}
dx\,S_0'(x)/\eta
= 2\pi (\ell_k+1/2)\,,
\label{SL2R-WKB}
\ee
with the ``action'' differential, $dx\,S_0'(x)$, and the Riemann surface,
$\Gamma_N$, the same as in the $SL(2,\mathbb{C})$ case, Eqs.~\re{dS_0} and
\re{curve}, respectively. A crucial difference between the $SL(2,\mathbb{C})$ and
$SL(2,\mathbb{R})$ magnets is that the integrals of motion, $q_n$, and the
separated variables, $x$ and $p_x$, take real values in the latter case. As a
consequence, the classical motion in the separated $SL(2,\mathbb{R})$ coordinates
is restricted to the intervals on the real $x-$axis on which $t_N^2(x)-4x^{2N}\ge
0$ (see Eq.~\re{w}). The integration in \re{SL2R-WKB} goes along the cycles (real
slices of $\Gamma_N$) encircling these intervals on the complex $x-$plane. In the
$SL(2,\mathbb{C})$ case, the classical trajectories wrap arbitrarily around
$\Gamma_N$ and, as a consequence, the quantization conditions \re{quan-cond}
involve also $\beta-$cycles which correspond, from point of view of the
$SL(2,\mathbb{R})$ magnet, to classically forbidden zones.

We shall solve the quantization conditions \re{quan-cond} in Section~5.

\section{Quasiclassical spectrum}

The WKB analysis performed in the previous Section allowed us to formulate the
quantization conditions for the integrals of motion, Eqs.~\re{quan-cond}, and
construct the WKB expression for the wave function in the separated coordinates,
Eq.~\re{Q-gen}. Let us extend our analysis and evaluate the physical observables
of the model -- the energy, $E_N$, and the quasimomentum, $\theta_N$.

Our starting point will be the expressions for $E_N$ and $\theta_N$ obtained in
Ref.~\cite{DKM-I,DKKM} within the method of the Baxter $\mathbb{Q}-$operator (see
Eqs.~\re{Energy} and \re{Quasimomentum} below). They are formulated in terms of
the eigenvalue of the Baxter $\mathbb{Q}-$operator, $Q(u,\bar u)$, which is a
function of two complex spectral parameters, $u$ and $\bar u$, such that
$i(u-\bar u)=n$ with $n$ arbitrary integer. The wave function in the SoV
representation, $Q(x,\bar x)$ (see Eq.~\re{SoV}), coincides with this function
for $u=\nu-in/2$ and $\bar u=\nu-in/2$ with $\nu$ real. In this way, $Q(u,\bar
u)$ can be considered as an analytical continuation of the wave function from the
real axis to the whole complex $\nu-$plane.

The energy spectrum of the quantum $SL(2,\mathbb{C})$ magnet is related to the
behaviour of the function $Q(u,\bar u)$ around two special points on the complex
$\nu-$plane corresponding to $u=\pm is$ and $\bar u=\pm i\bar s$ with $(s,\bar
s)$ being a single-particle $SL(2,\mathbb{C})$ spin, Eq.~\re{spins}. We notice
that this behaviour can not be deduced from the obtained WKB expression for the
$Q-$function \re{Q-gen} since the latter is valid only for large $u$ and $\bar
u$. In this Section, we shall construct an asymptotic expression for $Q(u,\bar
u)$, valid for large charges $q_n$ and $u,\bar u=\rm fixed$, and use it to
calculate the quasiclassical energy spectrum.

\subsection{Baxter $Q-$blocks}

To begin with, we summarize, following \cite{DKKM}, the main properties of the
eigenvalues of the Baxter ${\mathbb Q}-$operator, $Q(u,\bar u)$. The function
$Q(u,\bar u)$ can be decomposed into a bilinear combination of chiral blocks
\be
Q(u,\bar u)= \e^{-i\delta} Q_0(u) \widebar Q_0(\bar u) - \e^{i\delta} \, Q_1(u)
\widebar Q_1(\bar u)\,,
\label{Q-dec}
\ee
with $\delta$ arbitrary complex. The blocks $Q_0(u)$ and $Q_1(u)$ satisfy the
chiral Baxter equation \re{Bax-eq} for arbitrary complex $u$ and fulfil the
Wronskian condition
\be
Q_0(u+i)Q_1(u)-Q_0(u)Q_1(u+i)=\frac{\Gamma^N(iu-s)}{\Gamma^N(iu+s)}\,.
\label{Wron}
\ee
Similar blocks in the antiholomorphic sector, $\widebar Q_0(\bar u)$ and
$\widebar Q_1(\bar u)$, are defined as
\be
Q_1(u)=\frac{\Gamma^N(1-s+iu)}{\Gamma^N(s+iu)}\lr{\widebar Q_0(u^*)}^*\,,\qquad
\widebar Q_1(\bar u)=\frac{\Gamma^N(1-\bar s-i\bar u)}{\Gamma^N(\bar s-i\bar u)}
\lr{Q_0(\bar u^*)}^*\,.
\label{Q1}
\ee
The $Q-$blocks are meromorphic functions on the complex plane. Their analytical
properties can be summarized as
\be
Q_0(u)=\Gamma^{N-1}(1-s+iu) f(u)\,,\qquad
\widebar Q_0(\bar u)=\Gamma^{N-1}(1-\bar s-i\bar u) \bar f(\bar u)\,,
\label{Q0-poles}
\ee
where $f(u)$ and $\bar f(\bar u)$ are entire functions. As was already mentioned,
at $u=x$ and $\bar u=x^*$ with $x$ given by \re{x}, $Q(u,\bar u)$ defines the
wave function in the separated coordinates, Eq.~\re{SoV}. The constant $\delta$
in \re{Q-dec} is fixed by the requirement
%
that $Q(\nu-in/2,\nu+in/2)$ should not have poles for real $\nu$. The same
condition can be expressed as~\cite{DKKM}
\be
Q(i(1-s)+\epsilon,-i\bar s+\epsilon) = {\cal O}(\epsilon^0)\,,\qquad
Q(-is+\epsilon,i(1-\bar s)+\epsilon) = {\cal O}(\epsilon^0)\,,
\label{Q-finite}
\ee
for $\epsilon\to 0$ and $\bar s=1-s^*$.

The energy and the quasimomentum of the model are expressed in terms of the
$Q-$blocks as
\ba
E_N&=&-2\Im \lr{\ln Q_0(is)}'+2\Im \lr{\ln\widebar Q_0(-i\bar
s)}'+\varepsilon_N\,,
\label{Energy}
\\
\theta_N&=&i\ln\frac{Q_0(is)\,\widebar Q_0(i\bar s)}{Q_0(-is)\,\widebar Q_0(-i\bar
s)}\,.
\label{Quasimomentum}
\ea
where $\varepsilon_N=2N\Re\left[\psi(2s)+\psi(2-2s)-2\psi(1)\right]$.

Thus, the problem of calculating the energy spectrum of the model is reduced to
finding the blocks $Q_0(u)$ and $\widebar Q_0(\bar u)$. Their exact expressions
were obtained in \cite{DKKM}. In this Section we shall obtain asymptotic
expressions for the blocks $Q_0(u)$ and $\widebar Q_0(\bar u)$ valid for large
charges $q_n$ and fixed spectral parameters $u$ and $\bar u$.

\subsection{Asymptotic solutions to the Baxter equation}

Let us rewrite the Baxter equation \re{Bax-eq} for the block $Q_0(u)$ as
\be
(u+is)^N q(u) + (u-is)^N\frac1{q(u-i)}=t_N(u)\,,
\label{Bax-red}
\ee
where the notation was introduced for $q(u)= {Q_0(u+i)}/{Q_0(u)}$. We notice that
for large charges $q_n$ and fixed $u$ the polynomial $t_N(u)$, defined in
\re{curve}, takes large values $|t_N(u)|\gg 1$. This suggests to expand the
solutions to \re{Bax-red} in inverse powers of $t_N(u)$. Assuming that one of the
terms in the l.h.s.\ of \re{Bax-red} is much smaller than another one, we obtain
two solutions
\be
q_+(u)=\frac{t_N(u)}{(u+is)^N}+ \ldots \,,\qquad
q_-(u-i)=\frac{(u-is)^N}{t_N(u)}+ \ldots\,.
\label{q-red}
\ee
Here, ellipses denote subleading corrections controlled by the parameter
\be
\left|\frac{(u\pm is)^N(u\pm i(1-s))^N}{t_N(u)t_N(u\pm i)}
\right| \ll 1\,.
\label{ksi}
\ee
Eq.~\re{q-red} gives rise to two linear independent asymptotic solutions for the
block $Q_0(u)$. The general expression for $Q_0(u)$ with the prescribed
analytical properties \re{Q0-poles} takes the form
\be
Q_0^{\rm (as)}(u)=\frac{\Gamma^{N-1}(1-s+iu)}{\Gamma(s-iu)}
\varphi_+(u)+\frac1{\Gamma^N(s+iu)}\varphi_-(u)\,,
\label{Q0-as}
\ee
where $\varphi_\pm(u)$ are {\it entire\/} functions satisfying the functional
equations
\be
\frac{\varphi_+(u+i)}{\varphi_+(u)}=-i^N t_N(u)\,,\qquad
\frac{\varphi_-(u-i)}{\varphi_-(u)}=i^N t_N(u)\,.
\label{phi}
\ee
It is important to realize that the particular form of the ratio of the
$\Gamma-$functions in the r.h.s.\ of \re{Q0-as} is uniquely fixed by analytical
properties of the block $Q_0(u)$, Eq.~\re{Q0-poles}. If
%
any $\Gamma-$function in \re{Q0-as} was substituted as
$\Gamma(x)\to1/\Gamma(1-x)$, one would obtain another solution to
%
the Baxter equation \re{Bax-red} (up to
corrections suppressed by the factor \re{ksi}) but its pole structure would be
different from \re{Q0-poles}.

The $Q-$block in the antiholomorphic sector is given by similar expression
\be
\widebar Q_0^{\rm (as)}(\bar u)=\frac{\Gamma^{N-1}(1-\bar s-i\bar u)}
{\Gamma(\bar s+i\bar u)} \bar
\varphi_+(\bar u)+\frac1{\Gamma^N(\bar s-i\bar u)}\bar \varphi_-(\bar u)\,,
\label{Q0bar-as}
\ee
with $\bar\varphi_\pm(\bar u)$ entire functions satisfying the relations
\be
\frac{\bar\varphi_+(\bar u-i)}{\bar\varphi_+(\bar u)}=-(-i)^N \bar t_N(\bar u)\,,
\qquad
\frac{\bar\varphi_-(\bar u+i)}{\bar\varphi_-(\bar u)}={(-i)^N} \bar t_N(\bar
u)\,.
\label{phibar}
\ee
Here $\bar t_N(\bar u)$ is given by \re{curve} with the charges $q_n$ replaced by
their antiholomorphic counterparts $\bar q_n=q_n^*$, so that $\bar{t}_N(\bar
u)=t_N(\bar u^*)^*$. Substituting \re{Q0-as} and \re{Q0bar-as} into \re{Q1} we
obtain
\ba
Q_1^{\rm (as)}(u)&=&\frac{\sin(\pi(s+iu))}{\pi} {\Gamma^{N}(1-s+iu)}
\lr{\bar\varphi_+(u^*)}^*+\frac1{\Gamma^N(s+iu)}\lr{\bar\varphi_-(u^*)}^*\,,
\nonumber
\\
\widebar Q_1^{\rm (as)}(\bar u)&=&\frac{\sin(\pi(\bar s-i\bar u))}{\pi}
{\Gamma^{N}(1-\bar s-i\bar u)}\lr{\varphi_+(\bar u^*)}^*+\frac1{\Gamma^N(\bar
s-i\bar u)}
\lr{\varphi_-(\bar u^*)}^*\,.
\label{Q1-as}
\ea
To solve \re{phi} and \re{phibar} one factorizes the polynomial $t_N(u)$,
Eq.~\re{curve}, as
%
\be
t_N(u)=2u^N+q_2 u^{N-2} + ... + q_N=2 \prod_{k=1}^N (u-\lambda_k)\,.
\label{roots}
\ee
For arbitrary, large $q_n\sim 1/\eta^n$ its roots take, in general, large complex
values, $\lambda_n \sim 1/\eta$ and satisfy the sum rules
\be
\sum_k \lambda_k=0\,,\quad \sum_{k>n} \lambda_k\lambda_n = q_2/2\,,\quad \ldots
\,,\quad \prod_k \lambda_k =(-1)^N q_N/2\,.
\label{roots-1}
\ee
Substituting \re{roots} into \re{phi} one can write a particular solution for
$\varphi_+(u)$ in the form
\be
\varphi_+^{\rm (naive)}(u) \sim \e^{\pm \pi u}\,2^{-iu}\prod_{k=1}^N
\Gamma(i\lambda_k - iu)\,.
\label{naive}
\ee
As before, substituting $\Gamma(x)\to 1/\Gamma(1-x)$ in the r.h.s.\ of \re{naive}
one can get yet another solution to \re{phi}. To fix this ambiguity we require
that $\varphi_+(u)$ has to be an entire function of $u$ in the region
$u\sim\eta^0$. Therefore, the product of the $\Gamma-$function in the r.h.s.\ of
\re{naive} should not generate poles for $u\sim\eta^0$. Decomposing $\lambda_k$
into real and imaginary parts,
$\Gamma(i\lambda_k-iu)=\Gamma(i\Re\lambda_k-\Im\lambda_k-iu)$, one finds that in
spite of the fact that $\lambda_k\sim 1/\eta$, the roots with $\Im
\lambda_k\sim 1/\eta$ and $\Re\lambda_k\sim\eta^0$ generate the sequence of poles
located at $iu=i\Re\lambda_k-[\Im\lambda_k]+n\sim \eta^0$ with $n$ nonnegative
integer and $[...]$ denoting the entire part.

This suggests to separate all roots in \re{naive} into two sets according to $\Im
\lambda_k\ge 0$ and $\Im\lambda_k<0$ and look for the solutions to \re{phi} in the
form
\be
\varphi_+(u) = a_+(u)\,\widehat\varphi_+(u)\,,\qquad
\varphi_-(u) = a_-(u)\,\widehat\varphi_-(u)\,,
\label{phi-final}
\ee
where the notation was introduced for the basis functions
\ba
\widehat\varphi_+(u)&=& 2^{-iu}\prod_{\Im \lambda_k<0}
\Gamma(i\lambda_k - iu)\prod_{\Im \lambda_k\ge 0}
\frac1{\Gamma(1-i\lambda_k + iu)}\,,
\nonumber
\\
\widehat\varphi_-(u)&=&
 2^{iu}~\prod_{\Im \lambda_k\ge 0}
\Gamma(-i\lambda_k + iu)\prod_{\Im \lambda_k<0}
\frac1{\Gamma(1+i\lambda_k - iu)}\,.
\label{phi-minus-final}
\ea
In these expressions, the roots with $\Re\lambda_k\sim\eta^0$ do not generate
spurious poles.
In similar manner, the solutions to the antiholomorphic relations in \re{phibar}
are given by
\be
\widebar\varphi_+(\bar u)= \widebar a_+(\bar u)\, \lr{\widehat\varphi_+({\bar
u}^*)}^*\,,\qquad
\widebar\varphi_-(\bar u)= \widebar a_-(\bar u)\, \lr{\widehat\varphi_-({\bar
u}^*)}^*\,.
\label{phibar-final}
\ee
Substituting \re{phi-final} and \re{phibar-final} into \re{phi} and \re{phibar},
respectively, we find that the functions $a_\pm(u)$ and $\widebar a_\pm(\bar u)$
are entire (anti)periodic functions satisfying the functional relations
\be
a_\pm(u\pm i)=\mp(-1)^{N_-}a_\pm (u)\,,\qquad \widebar a_\pm(\bar u\mp
i)=\mp(-1)^{N_-} \widebar a_\pm (\bar u)\,.
\label{A-period}
\ee
Here, $N_-$ denotes the number of roots \re{roots-1} with  $\Im
\lambda_k$ negative
\be
N_-=\sum_{\Im \lambda_k<0}1\,,\qquad N_+=\sum_{\Im \lambda_k\ge  0}1\,,\qquad
N_++N_-=N\,.
\label{N_pm}
\ee
Inserting \re{phi-final} and \re{phibar-final} into \re{Q0-as} and \re{Q0bar-as},
one obtains asymptotic expressions for the Baxter blocks $Q_0(u)$ and $\widebar
Q_0(\bar u)$. They depend however on four yet undefined functions $a_\pm(u)$ and
$\widebar a_\pm(\bar u)$. Additional constraints on these functions are derived
in Appendix~\ref{App:match} (see Eq.~\re{AA-ratio}). They come from two different
sources. First, one has to ensure that the obtained expressions for the blocks
verify the Wronskian condition \re{Wron}. Second, for $u=\nu-in/2$ and $\bar
u=u^*$ (with $\nu$ real and $n$ integer) the function $Q(u,\bar u)$ coincides
with the wave function in the separated coordinates. Substituting the asymptotic
expressions for the blocks into \re{Q-dec} and continuing the resulting
expression to the region of large $u$, one should be able to match it into
analogous expression for the WKB wave function \re{Q-gen}. The matching procedure
is performed in Appendix~\ref{App:match}.

\subsection{Energy spectrum}

According to \re{Energy} and \re{Quasimomentum}, the energy and quasimomentum are
related to the behaviour of the blocks $Q_0(u)$ and $\widebar Q_0(\bar u)$ around
the points $u=\pm is$ and $\bar u=\pm i \bar s$. Substituting the obtained
asymptotic expressions for the blocks, Eqs.~\re{Q0-as} and \re{Q0bar-as}, into
the expression for the quasimomentum, Eq.~\re{Quasimomentum}, one finds after
some calculation (see Appendix~\ref{App:aux} for detail)
\be
\theta_N=-2\Theta=-2\Re \int_{\gamma_{P_0^-P_0^+}}dx\,S_{0}'(x)
=-2\Re \int_{P_0^-}^{P_0^+}dx\,S_{0}'(x)\qquad ({\rm mod}~ 2\pi)\,.
\label{quasi-integral}
\ee
Here in the second relation we replaced the phase $\Theta$ by its expression
\re{Theta} and put $\eta=1$. The contour integral in the r.h.s.\ of
\re{quasi-integral} depends on the integrals of motion $q_n$. Since the possible
values of the quasimomentum are given by $\theta_N=2\pi
\ell/N$ with $\ell$ being integer, one should expect that for $q_n$
satisfying the quantization conditions \re{quan-cond} the integral takes the same
quantized values. Indeed, we demonstrate this property in Appendix~\ref{App:aux}
by an explicit calculation of \re{quasi-integral}.

The calculation of the energy \re{Energy} goes along the same lines (see
Appendix~\ref{App:aux}). It leads to the following asymptotic expression for the
energy
\ba
E_N^{\rm (as)}&=&4\ln 2 +2\Re \sum_{\Im\lambda_k\ge 0}
\bigg[\psi(1-s-i\lambda_k)+\psi(s-i\lambda_k)-2\psi(1)\bigg]
\nonumber
\\
&&\phantom{4\ln 2}+2\Re \sum_{\Im\lambda_k<0}
\bigg[\psi(1-s+i\lambda_k)+\psi(s+i\lambda_k)-2\psi(1)\bigg]\,.
\label{Energy-1}
\ea
Here, the sum goes over the (complex) roots of the polynomial $t_N(u)$,
Eqs.~\re{roots} and \re{roots-1}. By the definition, their total number equals
the number of particles, $N$, and their values depend on the integrals of motion
$q_n$. Since the latter are quantized according to \re{q_n}, the relation
\re{Energy-1} establishes the dependence of the energy on the total set of
quantum numbers, Eq.~\re{q_n}.

The obtained expression for the energy \re{Energy-1} is symmetric under $s\to
1-s$. This is in agreement with the fact that the $SL(2,\mathbb{C})$
representations of the spin $s$ and $1-s$ are unitary equivalent and, as a
consequence, the corresponding spin magnets should have the same energy spectrum.

According to \re{Energy-1}, the roots with $\Im\lambda_k>0$ and $\Im\lambda_k<0$
provide different contribution to the energy. To elucidate this property let us
examine \re{Energy-1} for the $SL(2,\mathbb{C})$ spin $s=0$
\ba
E_N^{\rm (as)}\bigg|_{s=0}=4\ln 2 +2\Re \sum_{k=0}^N
\bigg[\psi(1+i\Re\lambda_k+|\Im\lambda_k|)+\psi(i\Re\lambda_k+|\Im\lambda_k|)
-2\psi(1)\bigg]\,,
\label{E-spec}
\ea
where $\lambda_k$ are roots of the polynomial $t_N(u)$, Eq.~\re{roots}. We recall
that the Euler $\psi-$function has poles at nonpositive integer values of its
argument. In the r.h.s.\ of \re{E-spec} these poles are never approached for
arbitrary charges $q_n$ provided that $q_N\neq 0$.

\begin{figure}[th]
\psfrag{nu}[cc][cc]{$\nu_h$}
\psfrag{E3}[bc][cc]{$E_3$}
\psfrag{E4}[bc][cc]{$E_4$}
\centerline{{\epsfysize5.6cm \epsfbox{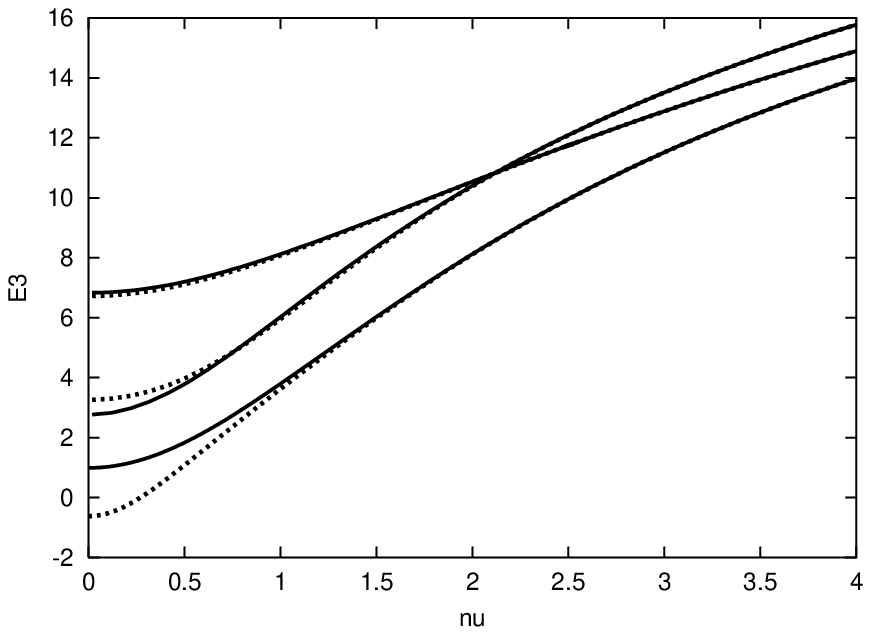}}\qquad{\epsfysize5.6cm \epsfbox{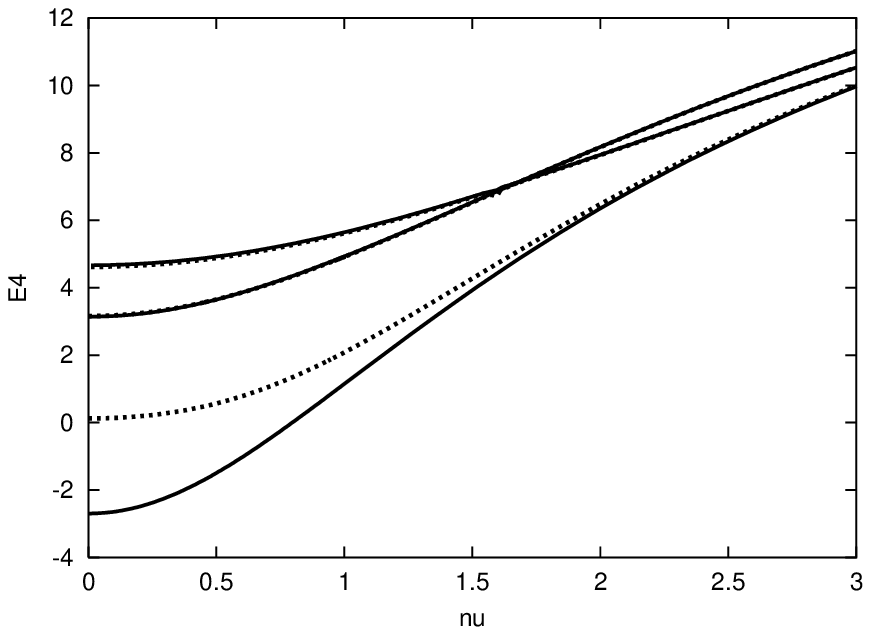}}}
\caption[]{Dependence of the energy $E_N^{(s=0)}$ on the total spin $h=1/2+i\nu_h$
for the three lowest eigenstates at $N=3$ (left panel) and $N=4$ (right panel).
The solid lines denote the exact energy from Refs.~\cite{KKM,DKKM}, the dotted
lines describe the asymptotic expression \re{E-spec} calculated for the exact
charges $q_3$ and $q_4$.}
\label{Fig:energy}
\end{figure}

Let us compare the asymptotic expression for the energy, Eq.~\re{E-spec}, with
the results of the exact calculation~\cite{KKM,DKKM}. For the sake of simplicity,
we choose the total Lorentz spin of the model to be $n_h=0$, so that the total
$SL(2,\mathbb{C})$ spin \re{h} equals $h=1/2+i\nu_h$, and examine the dependence
of the energy $E_N^{\rm (as)}|_{s=0}$ on real $\nu_h$ along a few lowest lying
trajectories at $N=3$ and $N=4$. Applying \re{E-spec} and \re{roots}, we
substitute $q_2=1/4+\nu_h^2$ and replace the charges $q_3,\ldots,q_N$ by their
exact values found in Ref.~\cite{KKM,DKKM}. Comparison with the exact expression
for the energy is shown in Figure~\ref{Fig:energy}. We recall that
Eqs.~\re{Energy-1} and \re{E-spec} were obtained to the leading order of the WKB
expansion under assumption that the ``energies'' $q_2,\ldots,q_N$ are large.
Therefore, it is not surprising that the quasiclassical expression for the ground
state trajectory agrees with the exact energy only for $\nu_h>1$. At the same
time, for the excited trajectories the agreement is rather remarkable (especially
at $N=4$) even for smaller $\nu_h$.

In Section~3.2 we already drew the analogy between the quantization conditions
for the $SL(2,\mathbb{C})$ and $SL(2,\mathbb{R})$ magnets. It can be further
extended to the asymptotic expressions for the energy. For the $SL(2,\mathbb{R})$
magnet the corresponding expression looks like~\cite{K2,BDKM,AB}
\be
SL(2,\mathbb{R}):\qquad E_N^{\rm (as)}\bigg|_{s=0}=2\ln 2 +\Re \sum_{k=0}^N
\bigg[\psi(1+i\lambda_k
)+\psi(i\lambda_k) -2\psi(1)\bigg]\,.
\label{E-spec-real}
\ee
Here, in distinction with the $SL(2,\mathbb{C})$ case, the roots $\lambda_k$ take
strictly real values. Similarity between Eqs.~\re{E-spec} and \re{E-spec-real}
suggests that there should exist a relation between the energy spectrum of the
$SL(2,\mathbb{C})$ and $SL(2,\mathbb{R})$ magnets. Indeed, it can be shown that
the latter can be obtained from the former by {\it analytical continuation\/} in
the total spin of the system, Eq.~\re{h}, from real $\nu_h$ to pure imaginary
$\nu_h$.

\section{Solving the quantization conditions}

In previous Section we derived the WKB quantization conditions for the integrals
of motions of the $SL(2,\mathbb{C})$ magnet, Eq.~\re{quan-cond}, and obtained the
expressions for the energy spectrum, Eqs.~\re{Energy-1} and \re{quasi-integral}.
In this Section, we shall solve the quantization conditions \re{quan-cond} and
reconstruct a fine structure of the spectrum.

The general solutions to \re{quan-cond} have the form \re{q_n}. The spectrum of
quantized charges $q_n$ is parameterized by the set of integers
$\Mybf{\ell}=(\ell_1,\ldots,\ell_{2N-4})$ entering the r.h.s.\ of \re{quan-cond},
as well as by continuous real $\nu_h$ and integer $n_h$ defining the total
$SL(2,\mathbb{C})$ spin of the magnet, Eq.~\re{h}. We recall that the
quantization conditions \re{quan-cond} involve the periods of the ``action''
differential over the canonical basis of oriented cycles,
$\Mybf{\alpha}=(\alpha_1,\ldots,\alpha_{N-2})$ and
$\Mybf{\beta}=(\beta_1,\ldots,\beta_{N-2})$, on the Riemann surface $\Gamma_N$ of
genus $g=N-2$. The definition of this basis on $\Gamma_N$ is ambiguous~\cite{D}
\be
 \Mybf{\alpha} \to \Mybf{\alpha'}=a\cdot\Mybf{\alpha} + b\cdot\Mybf{\beta}\,,\qquad
\Mybf{\beta} \to \Mybf{\beta'}= c\cdot \Mybf{\alpha} + d\cdot\Mybf{\beta}\,,
\label{cycle-trans}
\ee
where $a$, $b$, $c$ and $d$ are $(N-2)\times (N-2)$ matrices with integer entries
such that
\be
Z= \lr{\begin{array}{cc}
  a & b \\
  c & d
\end{array}}\,,\qquad \det Z=1\,,\qquad
Z^t\, {\small \lr{\!\!\!\begin{array}{rc}  0 & \II \\ -\II & 0 \end{array}}}\,
Z={\small \lr{\!\!\!\begin{array}{rc}  0 & \II \\ -\II & 0 \end{array}}}\,,
\ee
so that $Z\in {\rm Sp}(N-2,\mathbb{Z})$. Notice that $Z\in SL(2,\mathbb{Z})$ for
$N=3$. The spectrum of the model should not depend on the choice of the basis.
Indeed, the quantization conditions \re{quan-cond} stay invariant under the ${\rm
Sp}(N-2,\mathbb{Z})$ transformation \re{cycle-trans} provided that the integers
$\Mybf{\ell}_{\rm odd}=(\ell_1,\ldots,\ell_{2N-5})$ and $\Mybf{\ell}_{\rm
even}=(\ell_2,\ldots,\ell_{2N-4})$
 are transformed in the same way
\be
\Mybf{\ell}_{\rm odd} \to \Mybf{\ell}'_{\rm odd} =a\cdot\Mybf{\ell}_{\rm odd}
+ b\cdot\Mybf{\ell}_{\rm even} \,,\qquad
\Mybf{\ell}_{\rm
even}\to\Mybf{\ell}'_{\rm even}= c\cdot \Mybf{\ell}_{\rm odd}  +
d\cdot\Mybf{\ell}_{\rm even}\,.
\label{ell-trans}
\ee
This relation establishes the correspondence between two different solutions to
the quantization conditions labelled by the sets of integers $\Mybf{\ell}$ and
$\Mybf{\ell}'$.

\begin{figure}[th]
\psfrag{nu}[cc][cc]{$\nu_h$}
\psfrag{q(nu0)}[cc][cc]{$\Mybf{q}(\nu_0)$}
\psfrag{q(nu1)}[cc][cc]{$\Mybf{q}(\nu_1)$}
\centerline{{\epsfysize6cm \epsfbox{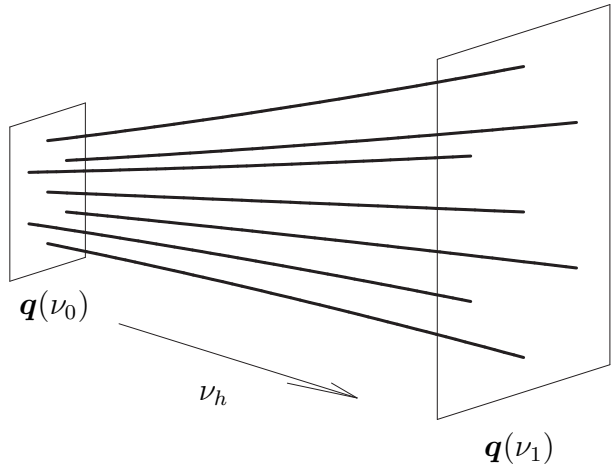}}}
\caption[]{Flow of the quantized values of the integrals of motion $\Mybf{q}=(q_2,\ldots,q_N)$ with
$\nu_h$.}
\label{Fig-flow}
\end{figure}

As was shown in Ref.~\cite{DKKM}, quantized values of the charges,
$q_n(\nu;n_h,\Mybf{\ell})$, form the family of one-dimensional nonintersecting
trajectories on the $(N-1)-$dimensional space of $\Mybf{q}=(q_2,...,q_N)$ (see
Figure~\ref{Fig-flow}). Each member of the family is labelled by integers $n_h$
and $\Mybf{\ell}$, while the ``proper time'' along the trajectory is defined by
$\nu_h$.
The quasiclassical approach 
allows us to reconstruct the trajectories in two different limits: ({\em i})
$\Mybf{\ell}={\rm large}$ and $\nu_h={\rm fixed}$; ({\em ii}) $\Mybf{\ell}={\rm
fixed}$ and $\nu_h={\rm large}$.
In the both cases, the integrals of motion take
large values, so that the quasiclassical approach is applicable. In the first
case, we shall find the points on the $\Mybf{q}-$space, at which the trajectories
pierce the hyperplane $\nu_h=\rm fixed$, and demonstrate that their positions
define a lattice-like structure on the moduli space (see Figure~\ref{Fig-flow}).
In the second case, we shall describe the flow of these points with $\nu_h$ along
the trajectories.

\subsection{Lattice structure}

Before we proceed with evaluating the contour integrals entering \re{quan-cond},
let us rewrite the quantization conditions in a slightly different form. >From the
expression for the quasimomentum \re{quasi-integral} one gets
\be
\theta_N=2\Re \int_{\gamma(\sigma_k)}
dx\,S_0'(x)=\frac{2\pi}{N}\ell \quad ({\rm mod}~ 2\pi)\,.
\label{quas-0}
\ee
Here the integration contour $\gamma(\sigma_k)$ goes from the point $P_0^+$ above
$x=0$ on the upper sheet of $\Gamma_N$ to the branching point $\sigma_k$,
encircles it and goes to the point $P_0^-$ on the lower sheet of $\Gamma_N$. The
quasimomentum $\theta_N$ describes the transformation properties of the
eigenfunction under the cyclic permutations of $N$ particles, $\Psi(\vec
z_2,\ldots,\vec z_N,\vec z_1)=\e^{i\theta_N}\Psi(\vec z_1,\ldots,\vec
z_{N-1},\vec z_N)$. Its value satisfy $\exp(iN\theta_N)=1$, so that $\ell$ is
integer in the r.h.s.\ of \re{quas-0}, $0\le\ell \le N-1$. One concludes from
\re{quas-0} that
\be
\frac1{\pi}\Re \int_{\gamma(\sigma_k)}dx\,S_0'(x) = \frac{\ell}{N}+n_k\,,
\label{Re-quas}
\ee
with $k=1,...,2(N-1)$ and $n_k$ integer. Remarkably enough, the quantization
conditions \re{quan-cond} are equivalent to the system of equations \re{Re-quas}.
To see this, one rewrites the $\alpha-$ and $\beta-$periods of the ``action''
differential as (see Figure~\ref{Fig-cuts})
$\oint_{\alpha_k}=\int_{\gamma(\sigma_{2k+1})}-\int_{\gamma(\sigma_{2k})}$ and
$\oint_{\beta_k}=\int_{\gamma(\sigma_{1})}-\int_{\gamma(\sigma_{2k})}$.
Substituting these relations into \re{quan-cond}, one finds that the quantization
conditions \re{quan-cond} can be expressed as a linear combination of integrals
entering the l.h.s.\ of \re{Re-quas}. In this way, one establishes the
correspondence between two sets of integers
\be
n_{2k+1}-n_{2k} =\ell_{2k-1}\,,\qquad n_{1}-n_{2k} =\ell_{2k}\,.
\label{n=l}
\ee
Eq.~\re{Re-quas} involves a rather complicated contour integral on the
hyperelliptic curve $\Gamma_N$. Although it is straightforward to calculate it
numerically for a given set of the integrals of motion $\Mybf{q}$, this does not
allow us to understand a general structure of solutions to \re{Re-quas}.

To this end, let us examine the quantization conditions \re{Re-quas} in the limit
\be
q_2=\mathcal{O}(\epsilon^0)\,,\qquad
q_n={\cal O}(\epsilon^{-(n-2)/2})\,,\qquad q_N={\cal O}(\epsilon^{-N/2})
\label{hierarchy}
\ee
with $\epsilon \ll 1$ and $n=3,...,N-1$. This hierarchy corresponds to large
$q_3,\ldots,q_N$ and fixed $q_2$.
The main advantage of \re{hierarchy} is that the spectral curve $\Gamma_N$,
Eq.~\re{curve}, simplifies significantly and integration in \re{quan-cond} can be
performed analytically. Indeed, after the scaling transformation  $x\to
(q_N/4)^{1/N}\, x$, $y\to q_N y$ the spectral curve \re{curve} takes the form
\be
\Gamma_N^{\rm (as)}: \qquad y^2=(x^N+1+p_{N-2}(x))\lr{1+p_{N-2}(x)} +{\cal O}\lr{\epsilon^2}\,,
\label{Gamma-as}
\ee
where
\be
p_{N-2}(x)=\sum_{n=2}^{N-1} u_n \,x^{N-n}\,,\qquad
u_n=\frac{q_n}4\lr{\frac{q_N}4}^{-n/N}={\cal O}(\epsilon)
\label{u-moduli}
\ee
Comparing \re{Gamma-as} with \re{curve} and \re{br_points}, we find that $N$
branching points of the curve $\Gamma_N^{\rm (as)}$ are located at the vertices
of the $N-$polygon, whereas the remaining $N-2$ points are located far from the
origin on the complex plane
\be
\sigma_m^{\rm (as)}=\e^{{i\pi}(2m-1)/N}\,,\qquad
\sigma_k^{\rm (as)}={\cal O}(\epsilon^{-1/(N-2)})\,,
\label{br_as}
\ee
where $m=1,...,N$ and $k=N+1,...,2(N-2)$.

It becomes straightforward to evaluate the hyperelliptic integral in \re{Re-quas}
for the first set of the branching points in \re{br_as} (see
Appendix~\ref{App:aux} for detail). Combining together \re{J-as}, \re{J-prop} and
\re{Re-quas} we find the system of $(N-2)$ equations for the integrals of motion
\ba
&&u_N \cdot N\,{\rm B}\lr{\frac12,\frac1{N}}- (u_2u_N)^*\cdot{\rm
B}\lr{\frac12,\frac{N-1}{N}}=\pi\sum_{k=1}^N \e^{-i\pi(2k-1)/N} n_k\,,
\nonumber
\\[2mm]
&&\lr{u_{N+1-m}u_N}\cdot {\rm B}\lr{\frac12,\frac{m}{N}}-
\lr{u_{m+1}u_N}^*\cdot{\rm B}\lr{\frac12,\frac{N-m}{N}}=\pi\sum_{k=1}^N
\e^{-i\pi(2k-1)m/N} n_k\,,
\label{Fourier}
\ea
with $m=2,...,N-2$. Here $B(x,y)=\Gamma(x)\Gamma(y)/\Gamma(x+y)$ is the Euler
beta-function, the moduli $u_m$ were defined in \re{u-moduli} for $2\le m\le N-1$
and $u_N=(q_N/4)^{1/N}$. One also gets the following relation for the
quasimomentum
\be
\ell=-\sum_{m=1}^{N} n_m \quad ({\rm mod}~ N)\,.
\label{quasi-n}
\ee
These relations were obtained in the small$-\epsilon$ limit and they hold up to
corrections $\sim\epsilon^{3/2}$. Notice that the sums in r.h.s.\ of
Eqs.~\re{Fourier} and \re{quasi-n} depend only on a subset of integers
$n_1,\ldots,n_{2N-4}$ corresponding to the first set of the branching points in
\re{br_as}. Comparing the same small$-\epsilon$ asymptotics of the the both sides
of \re{Fourier} one finds that $n_k\sim\epsilon^{-1/2}$. Then, Eq.~\re{n=l} leads
to $\ell_k\sim\epsilon^{-1/2}$.

Replacing in \re{Fourier} the moduli $u_n$ by their explicit expressions
\re{u-moduli}, one obtains the system of $(N-2)$ equations for the integrals of
motion $q_3,\ldots,q_N$. Since the second relation in \re{Fourier} is invariant
under $m\to N-m$, the number of independent relations reduces to $(N-1)$ real
equations and, therefore, the system \re{Fourier} is undetermined. The remaining
quantization conditions follow from the analysis of the integral \re{Re-quas} for
the second set of the branching points in Eq.~\re{br_as}. A straightforward
calculation shows that the corresponding $n-$integers, entering the r.h.s.\ of
\re{Re-quas}, scale in the limit \re{hierarchy} as $n_k\sim\epsilon^{1/2}$ and
induce subleading WKB corrections.

The quantized values of the ``highest'' charge $q_N$ can be calculated from the
first relation in \re{Fourier}. One finds after some algebra the following
remarkable expression
\be
q_N^{1/N}=\pi \frac{\Gamma(1+2/N)}{\Gamma^2(1/N)}\,\mathcal{Q}(\Mybf{n})
\left[1 + \frac{q_2^*}{\pi}
\frac{2N^2}{N-2}
{\cot({\pi}/{N})}
\left|\mathcal{Q}(\Mybf{n})\right|^{-2}+
\mathcal{O}\lr{{\left|\mathcal{Q}(\Mybf{n})\right|^{-4}}}
\right]\,,
\label{q_N-fin}
\ee
where the notation was introduced for the Fourier series
\be
\mathcal{Q}(\Mybf{n})=\sum_{k=1}^N
n_k\e^{-{i\pi}(2k-1)/N}\,,
\label{Four}
\ee
and $q_2^*=1/4+(\nu_h+in_h/2)^2-N(\nu_s+in_s/2)^2$ according to \re{h}.

The sum in the r.h.s.\ of \re{Four}
is invariant under simultaneous shift of integers, $n_k\to n_k+a$. This
transformation changes the value of $\ell$ in Eq.~\re{quasi-n}, but leaves
invariant the quasimomentum $\theta_N=\exp(2\pi i\ell/N)$. Therefore,
$q_N^{_{1/N}}$ and $\theta_N$ depend on the differences $n_k-n_{k+1}$, which in
their turn can be expressed in terms of the $\Mybf{\ell}-$integers with a help of
\re{n=l}. To obtain the corresponding expressions for $q_N^{_{1/N}}$ and
$\theta_N$ one chooses the ``gauge'' $n_1=0$ and substitutes in \re{Four} and
\re{quasi-n}
\be
n_{2k}= -\ell_{2k}\,,\qquad n_{2k+1}=
\ell_{2k-1}-\ell_{2k}\,,
\ee
with $k=1,...,N-2$. It follows from \re{q_N-fin} that, to the leading order of
the WKB expansion, $q_N^{1/N}$ does not depend on the total $SL(2,\mathbb{C})$
spin $h=(1+n_h)/2+i\nu_h$. The $h-$dependence enters into \re{q_N-fin} through
the nonleading correction, which becomes smaller as one goes to higher excited
states with larger $n_k$.

\subsubsection{Special case: $N=3$}

For $N=3$ we find from  \re{q_N-fin} and \re{quasi-n} the quantized values of the
charge $q_3$
\be
q_3^{1/3}=\frac{\Gamma^3(2/3)}{2\pi}\left[\frac12(n_1-2n_2+n_3)
+i\frac{\sqrt{3}}2(n_3-n_1)
\right]+\mathcal{O}(\epsilon^{1/2})\,,
\ee
and the quasimomentum $\ell=-n_1-n_2-n_3$, with $n_{1,2,3}\sim
\epsilon^{-1/2}$. Here, for simplicity we did not include the
nonleading correction $\sim q_2^*$. Using \re{n=l} one rewrites these relations
as%
\footnote{To make the correspondence with Ref.~\cite{DKKM}, one has
to redefine integers as $\ell_1+\ell_2\to \ell_1$ and $\ell_1-\ell_2\to \ell_2$.}
\ba
&q_3^{1/3}(\ell_1,\ell_2)&=\frac{\Gamma^3(2/3)}{2\pi}\left[\frac12(\ell_1+\ell_2)
+i\frac{\sqrt{3}}2(\ell_1-\ell_2)
\right]\,,
\nonumber
\\
&\theta_3(\ell_1,\ell_2)&=-\frac{2\pi}{3}(\ell_1+\ell_2)\qquad ({\rm
mod}~2\pi)\,.
\label{q3-N3}
\ea
Thus, to the leading order of the WKB expansion, the quantized charges
$q_3^{1/3}(\ell_1,\ell_2)$ are located on the complex plane at the vertices of
the lattice built from equilateral triangles.
\begin{figure}[th]
\psfrag{Im(q3^(1/3))}[cc][cc]{$\Im[q_3^{1/3}]$}
\psfrag{Re(q3^(1/3))}[cc][bc]{$\Re[q_3^{1/3}]$}
\centerline{{\epsfysize6cm \epsfbox{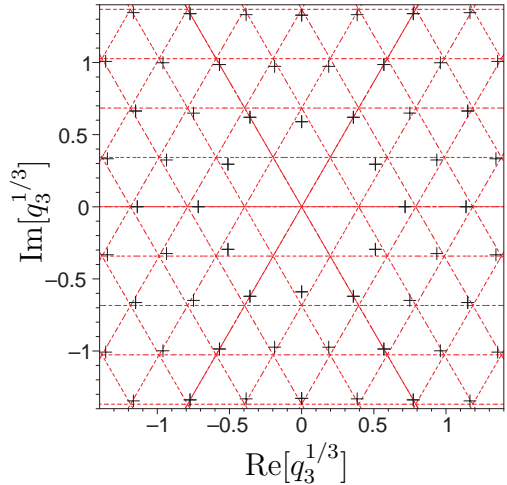}}}
\caption[]{Lattice structure at $N=3$. Crosses denote the exact values of $q_3^{1/3}$
at $q_2=1/4$. Dotted lines intersect at the points defined in Eq.~\re{q3-N3}.}
\label{Fig:q3}
\end{figure}

As one can see from Figure~\ref{Fig:q3}, Eq.~\re{q3-N3} is in a good agreement
with the exact results. Notice however that the exact values of $q_3$ do not
approach the origin, so that a few lattice vertices remain vacant. Since the
corresponding $q_3$ are small, one should not expect the WKB approach to be
applicable in this region. Indeed, one can verify that for $|\,q_3|<\Delta_3$,
the first nonleading WKB correction to the periods $a(q_3)$ and $a_D(q_3)$
becomes comparable with the leading order contribution.

\subsubsection{Special case: $N=4$}

For $N=4$ the spectrum of the magnet is parameterized by two quantum numbers
$q_3$ and $q_4$. From \re{q_N-fin} one gets the charge $q_4$ as
\be
q_4^{1/4}=\frac{\Gamma^2(3/4)}{4\sqrt{\pi}}\left[ \frac1{\sqrt {2}}
\lr{n_1-n_2-n_3+n_4}+
\frac{i}{\sqrt {2}}\lr{-n_1-n_2+n_3+n_4} \right]+\mathcal{O}(\epsilon^{1/2})\,,
\label{q4-quan}
\ee
where, in general, $n_{1,2,3,4}=\mathcal{O}(\epsilon^{-1/2})$. The quasimomentum
\re{quasi-n} is equal to
\be
\theta_4=-\frac{\pi}2\ell=\frac{\pi}2(n_1+n_2+n_3+n_4)\qquad ({\rm mod}~ 2\pi)\,.
\label{theta-4}
\ee
To find the charge $q_3$, we apply the second relation in \re{Fourier} at $N=4$,
$m=2$ and use the definition of the moduli \re{u-moduli}
\be
\Im \frac{q_3}{q_4^{1/2}} = (-n_1+n_2-n_3+n_4)+\mathcal{O}(\epsilon^{3/2})\,.
\label{q3-quan}
\ee
As was already explained, the system \re{Fourier} is undetermined and it does not
fix the charge $q_3$ completely. The additional relation on $q_3$ comes from the
analysis of the branching points in \re{br_as} located far from the origin.

The solutions to \re{q4-quan} and \re{q3-quan} are parameterized by three
integers $\ell_1=n_3-n_2$, $\ell_2=n_1-n_2$ and $\ell_4=n_1-n_4$. To reveal the
properties of the spectrum it is more convenient, however, to introduce their
linear combinations
\ba
m_1&=&\lr{n_1-n_2-n_3+n_4}/2=(n_1+n_4)+\frac{\ell}2\,,\qquad
\nonumber
\\
m_2&=&\lr{-n_1-n_2+n_3+n_4}/2=(n_3+n_4)+\frac{\ell}2\,,
\label{m's}
\ea
where integer $\ell$ defines the quasimomentum \re{theta-4}. Notice that
$m_{1,2}$ are integer for  $\ell=\rm even$ and half-integer for $\ell=\rm odd$.
Choosing the gauge $n_1=0$, one rewrites \re{q4-quan} and \re{q3-quan} as
\ba
&&q_4^{1/4}=\frac{\Gamma^2(3/4)}{2\sqrt{\pi}}\left( \frac{m_1}{\sqrt {2}}+
i\frac{m_2}{\sqrt {2}} \right)+\mathcal{O}(\epsilon^{1/2})\,,
\nonumber
\\[3mm]
&&\Im{\frac{q_3}{q_4^{1/2}}}=2\left(m_1-m_2-\frac{\ell}2\right)
+\mathcal{O}(\epsilon^{3/2})\,,
\label{q4-fin}
\ea
where $m_1=(-\ell_1+2\ell_2-\ell_4)/2$, $m_2=(\ell_1-\ell_4)/2$ and
$m_{1,2}=\mathcal{O}(\epsilon^{-1/2})$. Let us examine these expressions in more
detail. It is convenient to consider separately the $N=4$ eigenstates with
$q_3=0$ and $q_3\neq 0$.

For the eigenstates with $q_3=0$ one finds from \re{q3-quan} that
$n_1+n_3=n_2+n_4$. As a consequence, their quasimomentum,
$\theta_4=-\pi\ell/2=\pi(n_1+n_3)$, takes the values $\theta_4=0,\,\pi~({\rm
mod}~2\pi)$, while $m_1=n_1-n_2$ and $m_2=n_3-n_2$ are strictly integer. Thus,
the quantized values of $q_4$ for the eigenstates with $q_3=0$ are described by
the first relation in \re{q4-fin} with $m_{1,2}$ integer. They form a square
lattice on the complex $q_4^{_{1/4}}-$plane whose vertices are specified by the
pair of integers $(m_1,m_2)$. For $m_1\pm m_2$=even, $\ell=0$ (mod 4) and the
quasimomentum equals $\theta_4=0$. For $m_1\pm m_2$=odd, $\ell=2$ (mod 4) and the
quasimomentum equals $\theta_4=\pi$.

\begin{figure}[th]
\psfrag{Re(q4^(1/4))}[cc][bc]{$\Re[q_4^{1/4}]$}
\psfrag{Im(q4^(1/4))}[cc][cc]{$\Im[q_4^{1/4}]$}
\centerline{{\epsfysize6cm \epsfbox{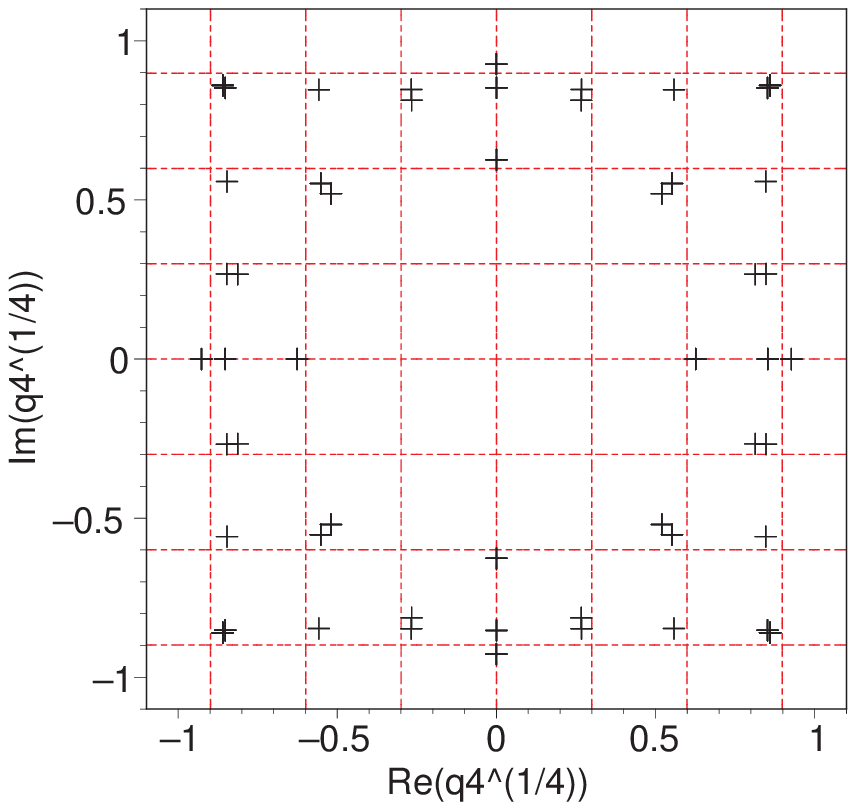}}\hspace*{20mm}
{\epsfysize6cm \epsfbox{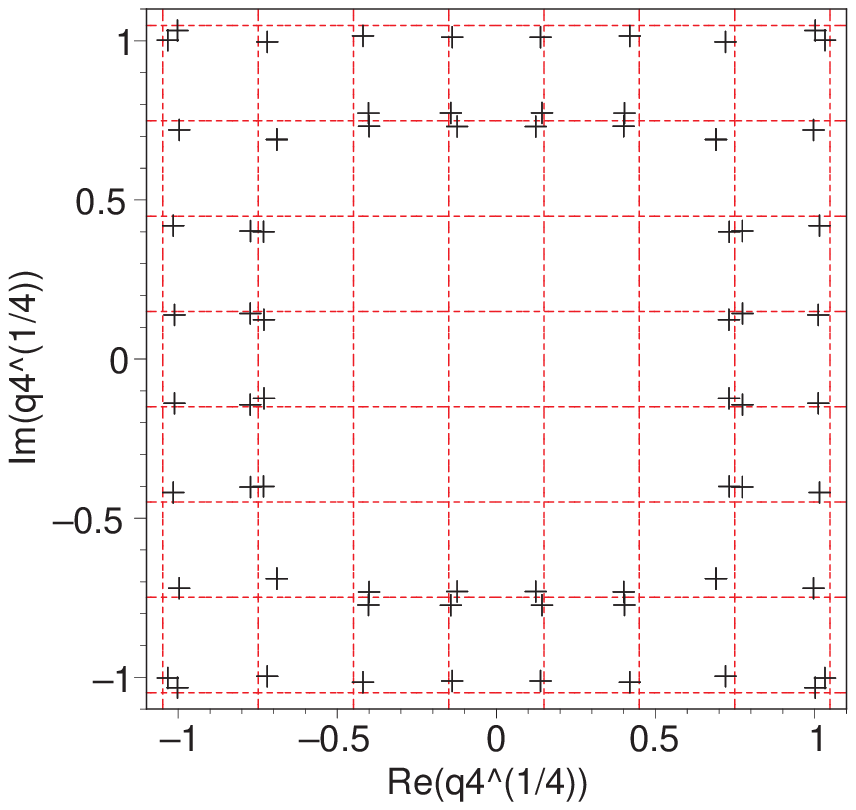}}}
\caption[]{Lattice structure at $N=4$. Crosses denote the exact values
of $q_4^{1/4}$ for different $q_3$ and $q_2=1/4$. Dotted lines intersect at the
points defined in Eq.~\re{q4-fin} with $m_{1,2}$ integer (left panel) and
$m_{1,2}$ half-integer (right panel).}
\label{Fig:q4}
\end{figure}

For $q_3\neq 0$ the eigenstates can be separated into two groups according to
their quasimomentum, $\theta_4=0,\,\pi$ and $\theta_4=\pm\pi/2$. Let us visualize
the solutions to \re{q4-fin} as points on the three-dimensional $\xi-$space with
the coordinates $\xi_1=\Re q_4^{1/4}$, $\xi_2=\Im q_4^{1/4}$ and
$\xi_3=\Im\lr{q_3/q_4^{_{1/2}}}$.
\begin{itemize}
\item $\theta_4=0,\,\pi$:
\quad
One finds from \re{m's} and \re{theta-4} that $\ell$ is even and $m_{1,2}$ are
{\it integer\/}. According to \re{q4-fin}, $q_4^{_{1/4}}$ does not depend on
$\ell$. Therefore, choosing $2(m_1-m_2)-\ell=0,\,\pm 2,\,\pm 4 ,...$, one finds
that the solutions to \re{q4-fin} define an infinite set of identical square
lattices in the $\xi-$space. These lattices run parallel to the
$(\xi_1,\xi_2)-$plane, as shown in Figure~\ref{Fig:q4} on the left, and cross the
$\xi_3-$axis at $\xi_3=0,\,\pm 2,\,\pm 4,...$. Exact results indicate
that the degeneracy between lattices with different $\xi_3$
is lifted by nonleading WKB corrections to $q_3$ and $q_4$.
\item $\theta_4=\pm\pi/2$:
\quad
One finds from \re{m's} and \re{theta-4} that $\ell$ is odd and $m_{1,2}$ are
{\it half-integer\/}. The solutions to \re{q4-fin} define an analogous lattice
structure in the $\xi-$space. The charges $q_4^{1/4}$ form square lattices on the
$(\xi_1,\xi_2)-$plane, shown in Figure~\ref{Fig:q4} on the right, which are dual
to the similar lattice in the previous case and have the coordinates $\xi_3=\pm
1,\,\pm 3,\,\pm 5,...$.
\end{itemize}
As follows from Figure~\ref{Fig:q4}, Eq.~\re{q4-fin} is in a good agreement
with the exact results of Refs.~\cite{KKM,DKKM}.%
\footnote{At the same time, Eq.~\re{q4-fin} invalidates the claim of Refs.~\cite{dVL-I,dVL-II}
that the charge $q_4$ may take only real values.} Surprisingly enough,
Eq.~\re{q4-fin} works throughout the whole spectrum including the ground state.
The latter is located at $q_3=0$ and $q_4$ given by \re{q4-fin} for $m_1=2$ and
$m_2=0$. Similar to the $N=3$ case, a few lattice sites remain unoccupied in the
small $q_4$ region, where the WKB approach is not applicable.

\begin{figure}[t]
\centerline{{\epsfysize6cm \epsfbox{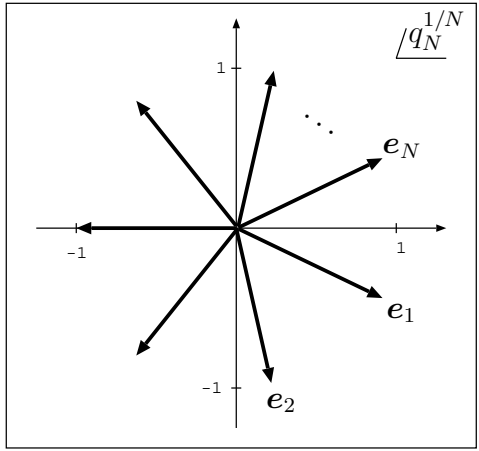}}}
\caption[]{Lattice structure on the complex $q_N^{1/N}-$plane is defined
by linear integer combinations of the vectors $\Mybf{e_1},\ldots,\Mybf{e_N}$.}
\label{Fig-lattice}
\end{figure}

Going over to higher $N$, we find that the lattice structure becomes more
complicated. The leading order expression for the ``highest'' charge,
Eq.~\re{q_N-fin}, can be written in the vector form
\be
q_N^{1/N}=\pi \frac{\Gamma(1+2/N)}{\Gamma^2(1/N)}
\cdot\sum_{k=1}^N
n_k\,\Mybf{e}_k+\mathcal{O}(\epsilon^{1/2})\,,
\label{e-lattice}
\ee
where $\Mybf{e}_k\equiv\e^{-{i\pi}(2k-1)/{N}}$ define unit vectors on the complex
plane as shown in Figure~\ref{Fig-lattice}. The quantized values of
$q_N^{_{1/N}}$ are given by linear combinations of these vectors with integer
(positive and negative) weights. Since $\sum_{k=1}^N\Mybf{e}_k=0$, the charge
$q_N^{1/N}$ depends on the differences $n_k-n_{k+1}$, or equivalently on the
$\Mybf{\ell}-$integers, Eq.~\re{n=l}.

\subsection{Whitham flow}

In the previous Section, we solved the quantization conditions \re{quan-cond} in
the limit \re{hierarchy}, which corresponds to large charges $q_3,\ldots,q_N$ and
fixed $q_2$, or equivalently the total $SL(2,\mathbb{C})$ spin
$h=(1+n_h)/2+i\nu_h$. Let us now determine the dependence of the charges \re{q_n}
on the continuous parameter $\nu_h$.

\begin{figure}[th]
\psfrag{nu}[tc][cc]{$\nu_h$}
\psfrag{ln_Im_q}[cc][cc]{$\ln{|\Im \widehat q_3|}$}
\psfrag{ln_Re_q}[cc][cc]{$\ln{|\Re \widehat q_3|}$}
\psfrag{O}[cc][cc]{$\textbf{\textsf{x}}$}
\centerline{{\epsfysize6cm \epsfbox{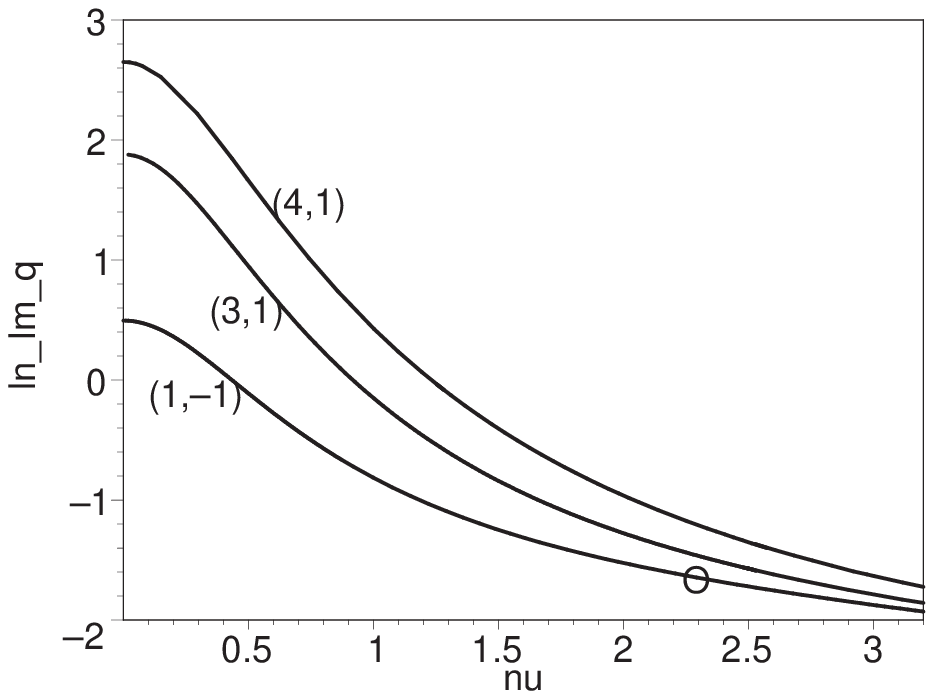}}\qquad {\epsfysize6cm \epsfbox{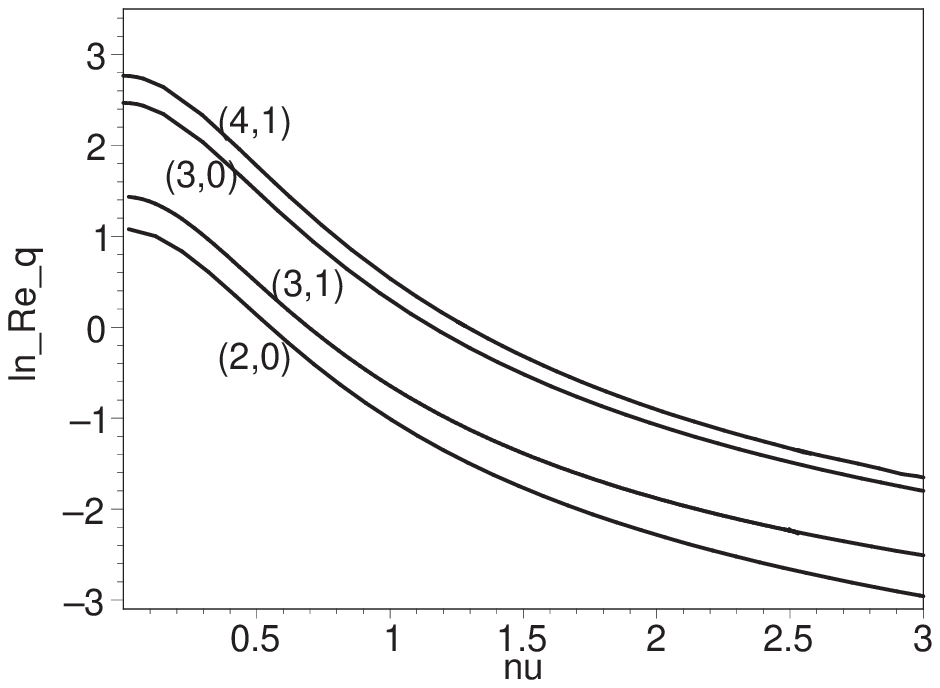}}}
\caption[]{The dependence of $\widehat q_3=q_3/q_2^{3/2}$ on the total spin $h=1/2+i\nu_h$ along
different trajectories specified by integers $(\ell_1,\ell_2)$. The charge
$\widehat q_3$ takes real values along the $(2,0)-$ and $(3,0)-$trajectories, and
pure imaginary values along the $(1,-1)-$trajectory. Cross denotes a point with
$\widehat q_3=\pm i/\sqrt{27}$. }
\label{Fig:Wh}
\end{figure}

To simplify analysis, we choose a single particle spin in \re{h} as $s=0$ and
consider the limit $\nu_h\gg n_h$. One finds from \re{h} that $q_2=1/4+\nu_h^2$
takes large positive values in this limit.%
\footnote{One can also consider the limit $n_h\gg \nu_h$, so that
$q_2=1/4-n_h^2/4$ is negative. Similar analysis allows one to find the
$n_h-$dependence of the quantum numbers $q_n$.} Let us explore an ambiguity in
choosing the WKB parameter $\eta$ in Eqs.~\re{t-hat} and \re{quan-cond} and fix
it as
\be
\eta=q_2^{-1/2}=(1/4+\nu_h^2)^{-1/2}\,,\qquad \widehat{q}_n=q_n q_2^{-n/2}\,,
\label{eta-value}
\ee
with $n=2,...,N$, so that $\widehat q_2=1$. In this Section, we shall solve the
quantization conditions \re{quan-cond} and determine the dependence of quantized
$\widehat{q}_n$ on $\eta$. As an example, we present in Figure~\ref{Fig:Wh} the
dependence $\widehat q_3=\widehat q_3(\nu_h;\ell_1,\ell_2)$ at $N=3$ for
different trajectories. We will show below that the $\nu_h-$dependence of the
charges is governed by the Whitham equations \cite{Wh}.

\subsubsection{Whitham equations}

Derivation of the Whitham equations is based on the following property of the
``action'' $S_0(x)$, Eq.~\re{dS_0},~\cite{IM}
\be
\frac{\partial}{\partial\eta} S_0(x)=\int_{x_0}^x  {dx}
\frac{\partial p_x}{\partial\eta}=\int_{x_0}^x \frac{dx}{y(x)}
\frac{\partial \widehat t_N(x)}{\partial\eta}=\sum_{k=3}^N
\frac{\partial \widehat q_k}{\partial\eta}\int_{x_0}^x
\frac{dx\,x^{N-k}}{y(x)}\,,
\label{ext-der}
\ee
where $\widehat t_N(x)=2x^N+x^{N-2}+\sum_{k=3}^N\widehat q_k\,x^{N-k}$ and $y(x)$
was defined in \re{w}. Differentiating the both sides of the first relation in
\re{quan-cond} and taking into account \re{ext-der}, one finds after some algebra
\be
\Re\left[\sum_{j=3}^N {\partial_\eta \widehat q_j}\cdot
\oint_{\alpha_k}\frac{dx\,x^{N-j}}{y(x)}\right]=\pi
\ell_{2k-1}\,.
\label{Wh}
\ee
The second equation in \re{quan-cond} leads to a similar relation with
$\alpha_k\to\beta_k$ and $\ell_{2k-1}\to\ell_{2k}$. The l.h.s.\ of \re{Wh}
involves a (unnormalized) differential of the first kind on $\Gamma_N$. Its
$\alpha-$ and $\beta-$periods can be parameterized as
\be
\oint_{\alpha_k}\frac{dx\,x^{j-1}}{y(x)}=2\pi \left[U^{-1}(\widehat q)\right]_{jk}\,,
\qquad
\oint_{\beta_k}\frac{dx\,x^{j-1}}{y(x)}=2\pi
\left[U^{-1}(\widehat q)\cdot\tau(\widehat q)\right]_{jk}\,,
\label{Wh-periods}
\ee
where $j,k=1,...,N-2$. Here $\tau=[\tau_{jk}(\widehat q)]$ is the Riemann matrix
for the hyperelliptic curve $\Gamma_N$, while the matrix $U=[U_{kj}(\widehat q)]$
defines the normalized holomorphic differentials (see Eq.~\re{exp-diff} in
Appendix~\ref{App:aux}). Both matrices depend on the charges $\widehat q_n$ and
are independent from the flow parameter $\eta$. In addition, the Riemann
$\tau-$matrix is symmetric and has positively definite imaginary part~\cite{D}.

To solve \re{Wh} one considers linear combinations
$X_k=\sum_{j=1}^{N-2}\partial_\eta\hat q_{N+1-j} \,[U^{-1}(\hat q)]_{jk}$. They
satisfy the relations
\be
2\Re X_k = \ell_{2k-1}\,,\qquad 2\Re
\sum_{j=1}^{N-2}X_j\,\tau_{jk}=\ell_{2k}\,,
\ee
whose solution can be written in the matrix form as
\be
\Mybf{X}(\widehat q)=-\frac{i}{2\Im \tau}\left(\Mybf{\ell}_{\rm even}-\tau^\dagger\cdot
\Mybf{\ell}_{\rm odd}\right)=-\left(\Mybf{\ell}_{\rm even}^{\,t}-
\Mybf{\ell}_{\rm odd}^{\,t}\cdot\tau^\dagger\right)\frac{i}{2\Im \tau}\,,
\label{X-sol}
\ee
where the vectors $\Mybf{\ell}_{\rm even}$ ($\Mybf{\ell}_{\rm odd}$) are built
from integers $l_{2k}$ ($\ell_{2k-1}$), and $\Im
\tau=(\tau-\tau^\dagger)/(2i)$ is a positively definite symmetric matrix.
Finally, one replaces $\Mybf{X}(\widehat q)$ in \re{X-sol} by its definition,
puts $\eta=q_2^{-1/2}$ and finds the system of Whitham equations
\be
 q_2^{3/2}\frac{\partial\,\widehat q_n}{\partial
 \,q_2}=\frac{i}4\left[\left(\Mybf{
\ell}_{\rm even}^{\,t}-
\Mybf{\ell}_{\rm odd}^{\,t}\cdot\tau^\dagger(\widehat q)\right)\frac{1}
{\Im \tau(\widehat q)}\cdot U(\widehat q)\right]_{N+1-n}
\label{Wh-flow-N}
\ee
with $n=3,...,N$. Notice that the r.h.s.\ of \re{Wh-flow-N} is $q_2-$independent
and it depends only on the charges $\widehat q_n$.

The Whitham equations \re{Wh-flow-N} involve the matrices $U(\widehat q)$ and
$\tau(\widehat q)$ defined in \re{Wh-periods}. Their explicit expressions depend
on the choice of the canonical set of the oriented cycles on $\Gamma_N$. As was
already mentioned, the $\alpha-$ and $\beta-$periods are defined up to
transformation \re{cycle-trans}, which acts on $U(\widehat q)$ and $\tau(\widehat
q)$ as
\be
U \to (a+b\tau)^{-1} U
\,,\qquad \tau\to (c+b\tau)(a+b\tau)^{-1}\,.
\label{U-trans}
\ee
It is easy to verify that the Whitham equations \re{Wh-flow-N} are invariant
under \re{U-trans} provided that $\Mybf{\ell}_{\rm odd}$ and $\Mybf{\ell}_{\rm
even}$ are transformed according to \re{ell-trans}.

The Whitham equations \re{Wh-flow-N} describe the dependence of the charges,
$q_n=q_n(\nu_h,\Mybf{\ell})$ on the total spin of the system $\nu_h$. They define
the flow of the quantum numbers $q_n$ with $\nu_h$ along the trajectory labelled
by the integers $\Mybf{\ell}$ (see Figure~\ref{Fig-flow}). To solve
\re{Wh-flow-N} one has to specify the initial conditions for $\widehat q_n$ (with
$n=3,...,N$) at some reference $q_2$. They are provided by the expressions for
the charges $q_n$, Eqs.~\re{q_N-fin} and \re{e-lattice}, obtained in the previous
section. We recall that Eqs.~\re{q_N-fin} and \re{e-lattice} were obtained in the
region of the moduli space \re{hierarchy} corresponding to $q_2=\rm fixed$. The
Whitham equations \re{Wh-flow-N} allow us to evolve the charges $q_n$ to
arbitrary large values of $q_2$.

\subsubsection{Whitham flow at $N=3$}

In the rest of this section, we shall present a detailed analysis of the Whitham
equations at $N=3$. Generalization to higher $N$ is straightforward and can be
performed along the same lines.

At $N=3$ it is convenient to introduce notation for the periods of the action
differential \re{dS_0}
\be
a(\widehat q_3)=\frac1{2\pi}\oint_{\alpha} dx\,\frac{(2x+3\widehat
q_3)}{y(x)}\,,\qquad a_D(\widehat q_3)=\frac1{2\pi}\oint_{\beta}
dx\,\frac{(2x+3\widehat q_3)}{y(x)}
\,.
\label{I-elliptic}
\ee
Here integration goes over the 
$\alpha-$ and $\beta-$cycles on the elliptic curve $\Gamma_3$, Eq.~\re{curve}
\be
\Gamma_3:\qquad y^2(x)=(x+\widehat q_3)(4x^3+x+\widehat q_3)
=4\prod_{j=1}^4(x-\sigma_j)\,.
\label{Gamma-3}
\ee
The genus of the Riemann surface defined by $\Gamma_3$ equals $g=N-2=1$. The
quantization conditions \re{quan-cond} for the charge $\widehat
q_3=q_3/q_2^{3/2}$ look like
\be
\Re a(\widehat q_3)=\frac{\ell_{1}}{2q_2^{1/2}}\,,\qquad
\Re a_D(\widehat q_3)=\frac{\ell_{2}}{2q_2^{1/2}}\,.
\label{a,a_D}
\ee
The Whitham equations \re{Wh-flow-N} take the following form at $N=3$
\be
q_2^{3/2}\frac{\partial\,\widehat q_3}{\partial \,q_2}=i
\frac{\ell_{2}-\ell_{1}\tau^*(\widehat q_3)}
{4\Im \tau(\widehat q_3)}\,U(\widehat q_3)\,,
\ee
with the functions $U(\widehat q_3)$ and $\tau(\widehat q_3)$ defined from
\re{Wh-periods} and \re{ext-der} as
\be
U(\widehat q_3)=\frac{1}{a'(\widehat q_3)}= 2\pi\lr{\oint_{\alpha}
\,\frac{dx}{y(x)}}^{-1}\,,\qquad
\tau(\widehat
q_3)=\frac{a_D'(\widehat q_3)}{a'(\widehat q_3)}
=\frac{\oint_{\beta}\,{dx}/{y(x)}}{\oint_{\alpha}\,{dx}/{y(x)}}\,,
\label{tau-3}
\ee
where $a'(\widehat q_3)=\partial a(\widehat q_3)/\partial \widehat q_3$ and
similar for $a_D'$. We recall that $\Im\tau(\widehat q_3)>0$ for arbitrary
$\widehat q_3$.

Performing integration in the r.h.s.\ of \re{I-elliptic}, one can evaluate
$a(\widehat q_3)$ and $a_D(\widehat q_3)$ in terms of the elliptic function of
the first and the second kinds. The resulting expressions for $a(\widehat q_3)$
and $a_D(\widehat q_3)$ are analytical functions on the complex $\widehat
q_3-$plane with two cuts. The cuts start at the values of $\widehat q_3$, for
which any two branching points of the curve $\Gamma_3$ merge,
$\sigma_j=\sigma_k$, and the integrand in \re{I-elliptic} develops a pole. This
happens for $q_3\to\infty$. The remaining singular points correspond to zeros of
the discriminant of the polynomial in the r.h.s.\ of \re{Gamma-3},
\be
16\prod_{j>k}(\sigma_j-\sigma_k)^2
=-\widehat q_3{}^6(1+27 \widehat q_3{}^2)\,.
\ee
In this way, one finds another three singular points on the complex $\widehat
q_3-$plane
\be
\widehat q_{3,\rm sing}=-\frac{i}{\sqrt{27}}\,,\quad 0\,,\quad \frac{i}{\sqrt{27}}\,.
\ee
Thus, the two cuts run on the complex $\widehat q_3-$plane between these three
points and $q_3=\infty$. As we will show below, the solutions to the quantization
conditions \re{a,a_D} can be obtained in a closed form at the vicinity of these
points.
Let us determine the asymptotic behaviour of the functions $a(\widehat q_3)$ and
$a_D(\widehat q_3)$ around the singular points, $\widehat q_3=0\,,-i/\sqrt{27}$
and $\infty$. The behaviour around $\widehat q_3=i/\sqrt{27}$ can be found by
making use of the symmetry of the curve \re{Gamma-3} under $x\to -x$ and
$\widehat q_3 \to -\widehat q_3$.

As the starting point, one has to specify the $\alpha-$ and $\beta-$cycles on the
curve $\Gamma_3$ (see Figure~\ref{Fig-cuts}). It is convenient to choose them in
such a way that the $\alpha-$ and $\beta-$cycles shrink into a point for
$\widehat q_3\to -i/\sqrt{27}$ and $\widehat q_3\to 0$, respectively. Obviously,
this choice is not unique and one can use another definition of the cycles. The
resulting expressions for the periods $a(\widehat q_3)$ and $a_D(\widehat q_3)$
are related to each other by the $SL(2,\mathbb{Z})$ transformation
\re{cycle-trans} and \re{ell-trans}.

At $q_3=0$ the branching points are located along the imaginary axis at
$\sigma_1=\sigma_2=0$, $\sigma_3=i/2$ and $\sigma_4=-i/2$. According to our
definition of the cycles, Figure~\ref{Fig-cuts}, the $\alpha-$cycle encircles the
cut $[\sigma_2,\sigma_3]$, whereas the $\beta-$cycle shrinks into a point.
Calculation of \re{I-elliptic} leads to
\be
a_D(0)=0\,,\qquad a(0)=\frac1{\pi}\oint_{\alpha}\frac{dx}{\sqrt{4x^2+1}}
=\frac{2}{\pi}\int^{i/2}_0\frac{dx}{\sqrt{4x^2+1}}=\frac{i}2\,.
\label{periods-0}
\ee
To obtain the behaviour of the functions $a(\widehat q_3)$ and $a_D(\widehat
q_3)$ in the vicinity of $\widehat q_3=0$, one examines their derivatives with
respect to $\widehat q_3$. According to \re{ext-der}, they are given by the
$\alpha-$ and $\beta-$periods of the holomorphic differential $dx/y(x)$ on
$\Gamma_3$, which can be calculated by the standard methods. The details of the
calculations can be found in \cite{K97}. In this way, one obtains the asymptotic
behaviour of the periods around $\widehat q_3=0$ as
\ba
&&a(\widehat q_3)=\frac{i}2-\frac{3}{\pi}\widehat q_3\left[\ln(i
\widehat q_3)-1\right]+...\,,
\nonumber
\\[3mm]
&&a_D(\widehat q_3)=i\widehat q_3+...\,,
\label{A-0}
\ea
where ellipses denote subleading $\mathcal{O}(\widehat q_3^2)-$terms.

At $\widehat q_3=-i/\sqrt{27}$ the branching points are located at
$\sigma_3=\sigma_2=i/\sqrt{12}$, $\sigma_1=i/\sqrt{27}$ and
$\sigma_4=-i/\sqrt{3}$. The $\beta-$cycle encircles $\sigma_2$ and $\sigma_1$,
while the $\alpha-$cycle shrinks into a point (see Figure~\ref{Fig-cuts}). The
periods \re{I-elliptic} are given by
\be
a(-i/\sqrt{27})=0\,,\qquad
a_D(-i/\sqrt{27})=\frac1{\pi}\int_{i/\sqrt{27}}^{i/\sqrt{12}}\frac{dx}
{\sqrt{(x+i/\sqrt{3})(x-i/\sqrt{27})}}=\frac{\ln 2}{\pi}\,.
\label{periods-1}
\ee
%
%
Expanding the periods in the vicinity of $\widehat q_3=-i/\sqrt{27}$ one finds
\ba
&&a(\widehat q_3)=\frac{i}3(1-\widehat q_3/q)
+...\,,\qquad
\nonumber
\\
&&a_D(\widehat q_3)=\frac{\ln 2}{\pi} +\frac{1}{6\pi}\lr{1-\widehat
q_3/q}\left[\,\ln\lr{1-\widehat q_3/q}-c\right]+...\,,
\label{A-1}
\ea
where $q=-i/\sqrt{27}$, $c=1+\ln(27/2)$ and ellipses denote
$\mathcal{O}((1-\widehat q_3/q)^2)-$terms.
%
%

For $\widehat q_3\to\infty$ the branching points $\sigma_k$ move away from the
origin and the spectral curve \re{Gamma-3} can be approximated as
$y^2=(x+\widehat q_3)(4x^3+\widehat q_3)$. Three branching points are located at
the vertices of equilateral triangle, $\sigma_k=(\widehat
q_3/4)^{1/3}\e^{i\pi(2k-1)/3}$ with $k=1,2,3$, and the last point at
$\sigma_4=-\widehat q_3$. The calculation of the periods \re{I-elliptic} leads to
\ba
a(\widehat q_3)&\stackrel{\widehat q_3\to\infty}{=}& 
{\widehat q_3^{\,1/3}}\frac{2\pi}{3\Gamma^3(2/3)}\cdot
\lr{\frac32-i\frac{\sqrt{3}}{2}}+...\,,
\nonumber
\\
a_D(\widehat q_3)&\stackrel{\widehat
q_3\to\infty}{=}&
\widehat
q_3^{\,1/3}\frac{2\pi}{3\Gamma^3(2/3)}\cdot\lr{\frac32+i\frac{\sqrt{3}}2}+...\,,
\label{a-large}
\ea
where ellipses denote subleading $\mathcal{O}(q_3^{-1/3})-$terms.
%
%
Notice that \re{a-large} can be written as $a(\widehat q_3)=(I_3-I_2)/(2\pi)$ and
$a_D(\widehat q_3)=(I_1-I_2)/(2\pi)$, where the integral $I_k$ was defined in
\re{J-int} for $N=3$.
We already encountered the same elliptic integral in Section~5.1, when we
analyzed the quantization condition in another region of the $\Mybf{q}-$space,
Eq.~\re{hierarchy}. Matching \re{eta-value} into \re{hierarchy} we find that
\be
\widehat q_3
={\cal O}(\epsilon^{-3/2})\,,\qquad
q_2
=\mathcal{O}(\epsilon^0)\,.
\ee
Thus, the two regions, Eqs.~\re{hierarchy} and \re{eta-value}, overlap as
$\widehat q_3\to\infty$ and $q_2=\rm fixed$. As a consequence, one expects that
at large $\widehat q_3$ the solutions to the quantization conditions \re{a,a_D}
should match the expressions for the quantum numbers $q_3$ obtained in the
previous Section, Eq.~\re{e-lattice}. Indeed, substitution of \re{a-large} into
\re{a,a_D} yields the expression
\be
\widehat q_3^{1/3}
=q_2^{-1/2}\frac{\Gamma^3(2/3)}{2\pi}
\left[\frac12 (\ell_1+\ell_2)
+i\frac{\sqrt{3}}2(\ell_1-\ell_2)
\right]+\mathcal{O}(q_2^{1/2})\,,
\label{q3-infinity}
\ee
which coincides with \re{q3-N3} since $\widehat q_3^{1/3}=q_3^{1/3} q_2^{-1/2}$.

Let us substitute the obtained expressions for the $a-$ and $a_D-$periods into
the quantization conditions \re{a,a_D} and derive the WKB expression for the
charge $q_3$. We remind that Eqs.~\re{A-0} and \re{A-1} hold in the vicinity of
$\widehat q_3=0$ and $\widehat q_3=-i/\sqrt{27}$, respectively.

For $|\widehat q_3| \ll 1$ one finds from \re{A-0} and \re{a,a_D}
\be
\Im \widehat q_3 =-\frac12\ell_2'\, q_2^{-1/2}\,,\qquad
\Re\left[\widehat q_3\lr{\ln(i  \widehat q_3)-1} \right]=-\frac{\pi}6\ell_1'\,
q_2^{-1/2}\,,
\label{q3-corr}
\ee
where $\widehat q_3=q_3/q_2^{3/2}$. Eq.~\re{q3-corr} defines the scaling
behaviour of quantized $q_3$ in the region $q_3\sim q_2$ for large $q_2$. Notice
that in comparison with \re{a,a_D} we replaced $(\ell_1,\ell_2)$ by another pair
of integers $(\ell_1',\ell_2')$. This was done in order to distinguish the
$\ell-$integers entering the r.h.s.\ of \re{q3-infinity} and \re{q3-corr}. As was
already mentioned, the periods $a(\widehat q_3)$ and $a_D(\widehat q_3)$ depend
on the definition of the $\alpha-$ and $\beta-$cycles on the Riemann surface
$\Gamma_3$. These cycles encircle the branching points $\sigma_k$ (see
Figure~\ref{Fig-cuts}), which are moved on the complex plane as $\widehat q_3$
varies. The two pairs of integers, $(\ell_1,\ell_2)$ and $(\ell_1',\ell_2')$,
would have been the same if, going from $\widehat q_3\to\infty$ to $\widehat
q_3=0$, we have traced the $\widehat q_3-$dependence of the $\alpha-$ and
$\beta-$cycles . Within our definition of the cycles, the two pairs are related
to each other by the $SL(2,\mathbb{Z})$ transformation \re{ell-trans}
\be
\ell_1'=-\ell_1-\ell_2\,,\qquad \ell_2'=-\ell_2\,.
\label{l-prime}
\ee
For $\widehat q_3\sim q=-i/\sqrt{27}$, one finds from \re{A-1} and \re{a,a_D}
\be
\Re\widehat q_3= \frac{1}{2\sqrt 3}  {\ell_1'}\,{q_2^{-1/2}}\,,\qquad
\Re\left(\lr{1-\widehat
q_3/q}\left[\,\ln\lr{1-\widehat q_3/q}-c\right]\right)=3\pi
{\ell_2'}\,{q_2^{-1/2}}-6\ln 2\,,
\label{q3-mid}
\ee
with $(\ell_1',\ell_2')$ the same as in \re{q3-corr} and \re{l-prime}. It follows
from this relation that $\Im\widehat q_3/\Re\widehat q_3 \sim \ln q_2$, so that
$\widehat q_3$ is dominated by its imaginary part at large $q_2$. Eq.~\re{q3-mid}
defines the scaling behaviour of quantized $q_3$ in the vicinity of the point on
the moduli space $q_3=-i(q_2/3)^{3/2}$.

For $\widehat q_3\to\infty$, the leading order expression for the charge is given
by \re{q3-infinity}. One can improve this relation by including nonleading
corrections to the periods in \re{A-1}. In this way, we obtain
\be
q_3^{1/3}=\frac{\Gamma^3(2/3)}{2\pi} \mathcal{Q}(\Mybf{n})
\left[1+\frac{3 \sqrt{3}}{2\pi}\frac{q_2}{|\mathcal{Q}(\Mybf{n})|^2}-
\lr{\frac{3 \sqrt{3}}{2\pi}\frac{q_2}{|\mathcal{Q}(\Mybf{n})|^2}}^2
+
\mathcal{O}(q_2^3)\right]\,,
\label{u-corr}
\ee
where
\be
\mathcal{Q}(\Mybf{n})=\frac12(\ell_1+\ell_2)+i\frac{\sqrt{3}}2(\ell_1-\ell_2)=\sum_{k=1}^3
n_k \,\e^{i\pi(2k-1)/3}\,.
\ee
One verifies that the first two terms in the r.h.s.\ of \re{u-corr} coincide with
\re{q_N-fin} for $N=3$ and $q_2$ real. Eq.~\re{u-corr} defines the scaling
behaviour of quantized $q_3$ in the region $q_3\gg q_2^{3/2}$.

By the construction, Eqs.~\re{q3-corr}, \re{q3-mid} and \re{u-corr} satisfy the
Whitham equation \re{Wh-flow-N} in the different regions of the
$(q_2,q_3)-$space. They define the $(\ell_1,\ell_2)-$trajectories, which start at
$q_2=1/4$ and go to larger $q_2=1/4+\nu_h^2$. Example of such trajectories is
shown in Figure~\ref{Fig:Wh}. In the regions $q_2<1$ and $q_2\gg 1$ the charge
$q_3$ satisfies Eqs.~\re{u-corr} and \re{q3-corr}, respectively. Solving the
Whitham equation \re{Wh-flow-N} and using \re{u-corr} as an initial condition at
$q_2=1/4$, one can reconstruct the flow of $q_3$ in the intermediate region
$q_2\sim 1$ along the trajectories labelled by integers $(\ell_1,\ell_2)$. Some
of these trajectories go in the complex $(q_2,q_3)-$space in the vicinity of the
point $q_3=-i(q_2/3)^{3/2}$. In particular, this is the case for the
$(1,-1)-$trajectory. The flow of $q_3$ around this point, shown by cross in
Figure~\ref{Fig:Wh} (see left panel), is described by Eq.~\re{q3-mid}.

\begin{figure}[th]
\psfrag{q2^(1/2)}[cc][rc]{$q_2^{1/2}$}
\psfrag{l1}[cc][cc]{$\ell_1^{\rm (as)}$}
\psfrag{l2}[cc][cc]{$\ell_2^{\rm (as)}$}
\centerline{{\epsfysize5.2cm \epsfbox{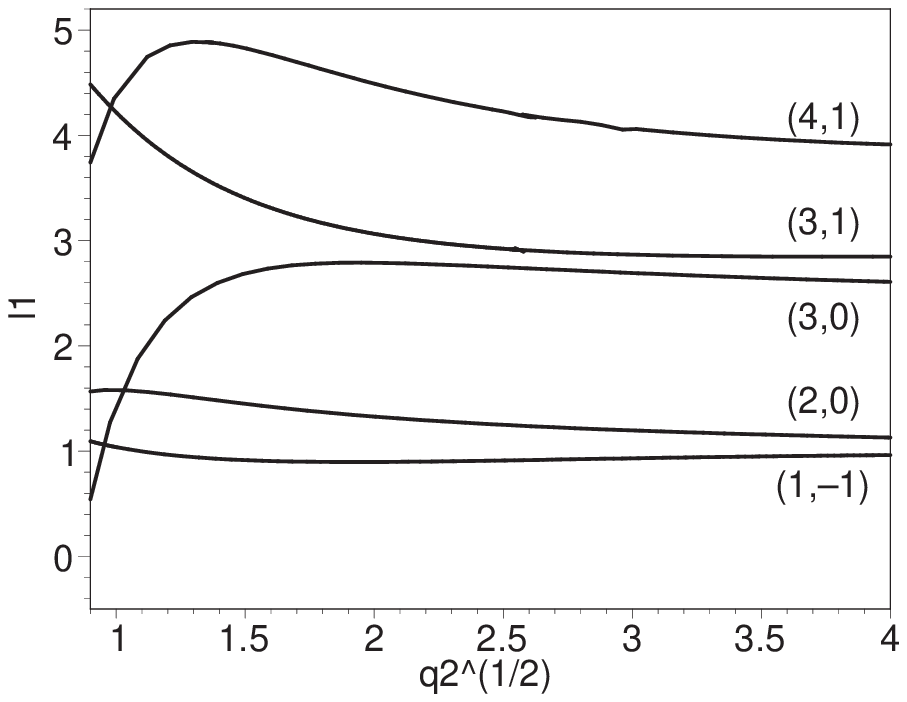}}\qquad
{\epsfysize5.3cm \epsfbox{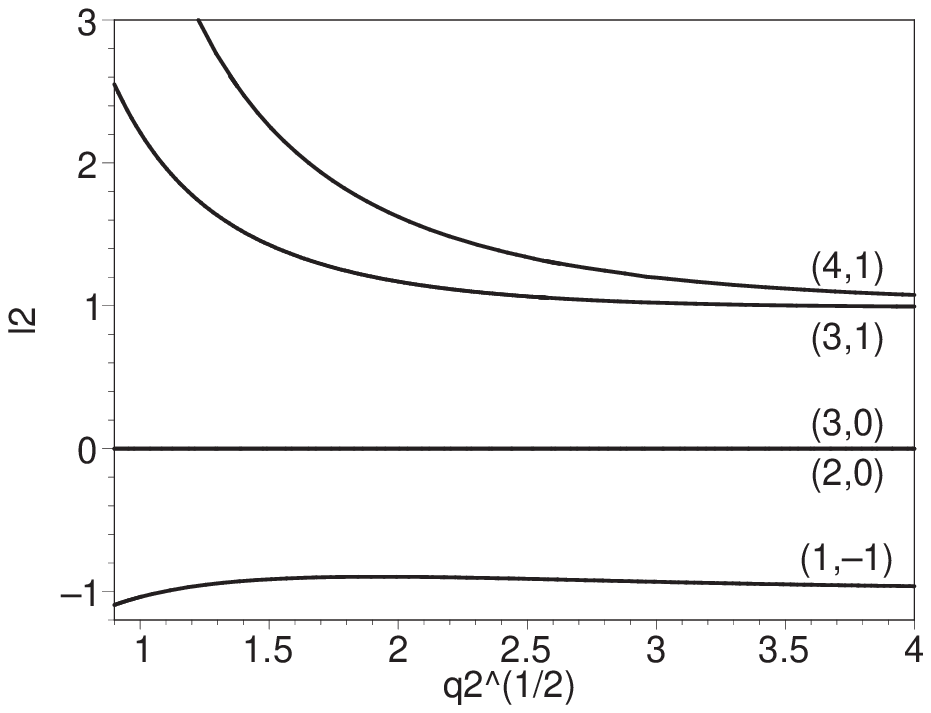}}}
\caption[]{
The functions
$\ell_1^{{\rm (as)}}=\ell_2'-\ell_1'$ (left panel) and $\ell_2^{{\rm
(as)}}=-\ell_2'$ (right panel) are calculated from \re{q3-corr} using the exact
eigenvalues $q_3$. Pairs of integers $(\ell_1,\ell_2)$ attached to the curves
specify different trajectories for the charge $q_3=q_3(\ell_1,\ell_2)$,
Eq.~\re{q3-N3}.}
\label{Fig-Whitham1}
\end{figure}
\begin{figure}[bh]
\psfrag{q2^(1/2)}[cr][cr]{$q_2^{1/2}$}
\psfrag{y3}[bc][cc]{$|q_3^{\rm (ex)}/q_3^{\rm (as)}|^{1/3}$}
\centerline{{\epsfysize5.2cm \epsfbox{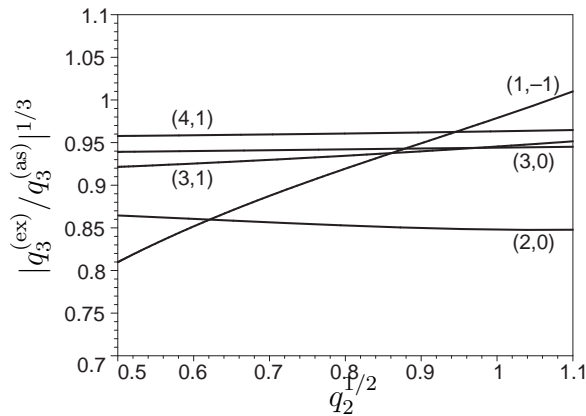}}}
\caption[]{Ratio of the exact, $q_3^{\rm (ex)}$, and the
asymptotic, $q_3^{\rm (as)}$, expressions for the charge $q_3$ at small $q_2$.
Different curves correspond to the same trajectories as in Figures~\ref{Fig:Wh}
and \ref{Fig-Whitham1}.}
\label{Fig-Whitham2}
\end{figure}
To verify the WKB quantization conditions, one substitutes the exact values of
$q_3$ into the l.h.s.\ of \re{q3-corr} and calculates the corresponding values of
$\ell_{1,2}$ for different $q_2$ using \re{l-prime}. In this way, for each
trajectory $q_3=q_3(\nu;\ell_1,\ell_2)$ we obtain two functions that we denote as
$\ell_{1,2}^{{\rm (as)}}(q_2)$. Few examples of such functions are shown in
Figure~\ref{Fig-Whitham1}. From \re{q3-corr} one would expect that for large
$q_2$ these functions should approach the same integer values $\ell_{1,2}$ as
those specifying the trajectories. Indeed, one finds from
Figure~\ref{Fig-Whitham1} that this happens for all trajectories except the
$(2,0)-$trajectory. In the latter case, the function $\ell_1^{{\rm (as)}}(q_2)$
approaches the value $1$ instead of expected $\ell_1=2$. The reason for this is
that the charge $\widehat q_3$ takes anomalously small values along the
$(2,0)-$trajectory (see Figure~\ref{Fig:Wh} on the right).  As a consequence, the
nonleading WKB correction to the ``action'' function in \re{WKB-exp} becomes
important for this particular trajectory. It provides a contribution to the
$a(\widehat q_3)$ and $a_D(\widehat q_3)$ comparable with the leading order
expression \re{q3-corr} and increases the value of $\ell_1^{{\rm (as)}}(q_2)$
improving an agreement with the exact result.

As can be seen from Figure~\ref{Fig-Whitham1}, Eq.~\re{q3-corr} is not satisfied
at small $q_2$. To describe the Whitham flow of the charge $q_3$ in this region,
one has to apply \re{u-corr}. Comparison of the exact $q_3$ with the asymptotic
expression \re{u-corr} for $q_2<1$ is shown in Figure~\ref{Fig-Whitham2}. One
observes that, aside from the $(2,0)-$trajectory, the agreement is rather good.
The accuracy can be further impoved by including nonleading WKB corrections in
\re{WKB-exp}.

Thus, the WKB quantization conditions, Eqs.~\re{q3-corr} and \re{u-corr},
successfully describe the exact spectrum of the charge $q_3$ shown in
Figure~\ref{Fig:Wh} for $1/4\le q_2<1$ and $q_2>1$, respectively.

\section{Conclusions}

In this paper, we have developed a quasiclassical approach to solving the
spectral problem for the noncompact $SL(2,\mathbb{C})$ Heisenberg spin magnet. It
allowed us to understand hidden symmetry properties of the energy spectrum. We
also demonstrated that the energy spectrum obtained within this approach is in a
good agreement with the exact results.

The model represents a generalization of the well-known spin$-1/2$ XXX chain to
infinite-dimensional representation of the $SL(2,\mathbb{C})$ group. Using
realization of spin operators as differential operators acting on the plane, one
can map the noncompact spin magnet of length $N$ into a two-dimensional
completely integrable quantum-mechanical model of $N$ particles with a
nearest-neighbour interaction. This model has appeared in high-energy QCD as
describing multi-gluonic compound states in the multi-colour limit. Due to its
complete integrability, the energy spectrum of the $N$ particle system is
uniquely specified by the total set of quantum numbers $q_2,\ldots,q_N$. The
latter are defined as eigenvalues of the mutually commuting integrals of motions
and their possible values are constrained by the quantization conditions.

Applying the methods of nonlinear WKB analysis~\cite{Wh}, we constructed the wave
function of the $N-$particle system in the representation of the separated
coordinates. To the leading order of the WKB expansion, the wave function is
determined by the ``action'' function, which satisfies the Hamiltonian-Jacobi
equations in the underlying classical model. In the classical case, the
noncompact magnet describes the system of interacting particles moving on the
two-dimensional plane. 
Solving the classical equations of motion, one finds that their collective motion
describes a propagation of the soliton wave in the closed chain of particles with
periodic boundary conditions. The same motion in the separated coordinates
corresponds to wrapping of classical trajectories around the Riemann surface
$\Gamma_N$ defined by spectral curve of the model. The charges $q_2,\ldots,q_N$
take arbitrary complex values in the classical model and define the moduli of
$\Gamma_N$.

The quantization conditions for the charges $q_2,\ldots,q_N$ follow from the
requirement for the wave function of the $N-$particle system to be a
single-valued function of the separated coordinates. To the leading order of the
WKB expansion, these conditions have the form of the Bohr-Sommerfeld relations
imposed on periodic orbits of the classical motion on the spectral curve
$\Gamma_N$. Solving the WKB quantization conditions, we demonstrated that, for
fixed total $SL(2,\mathbb{C})$ spin of the system, the eigenvalues of the
integrals of motion form a lattice structure on the moduli space of the model. At
$N=3$ and $N=4$ the lattices are built from equilateral triangles and squares,
respectively. The dependence of the charges on the total $SL(2,\mathbb{C})$ spin
is governed by the Whitham equations, which were solved at $N=3$ by making use of
the modular properties of the elliptic curve $\Gamma_3$.

A novel feature of the obtained quantization conditions is that they involve {\it
both\/} the $\alpha-$ and $\beta-$periods on $\Gamma_N$. Notice that in
conventional one-dimensional lattice integrable models, like the
$SL(2,\mathbb{R})$ Heisenberg spin magnet and the Toda chain model, the WKB
quantization conditions involve only the $\alpha-$cycles, since the
$\beta-$cycles correspond to classically forbidden zones. This implies that the
quantization conditions for the $SL(2,\mathbb{C})$ magnet are invariant under
modular transformations of the spectral curve. As a consequence, the energy
spectrum of the model possesses a hidden symmetry which is analogous to the
$S-$duality in the Yang-Mills theory~\cite{SW}.

In conclusion, we should mention that our consideration was restricted to the
leading order of the WKB expansion. The obtained expressions for the energy
spectrum can systematically improved by including nonleading WKB corrections.

\section*{Acknowledgements}

We are most grateful to A.~Gorsky and J.~Kota\'nski for collaboration at the
early stages of this project. We would like to thank R.~Janik, I.~Kogan,
F.~Smirnov and A.~Turbiner for useful discussions. This work was supported in
part by the grant 00-01-005-00 of the Russian Foundation for Fundamental Research
(A.M. and S.D.), by the Sofya Kovalevskaya programme of Alexander von Humboldt
Foundation (A.M.) and by the NATO Fellowship (A.M.).

\appendix
\renewcommand{\theequation}{\Alph{section}.\arabic{equation}}
\setcounter{table}{0}
\renewcommand{\thetable}{\Alph{table}}

\section{Appendix: Matching conditions}
\label{App:match}

Let us require that the obtained asymptotic expressions for the Baxter blocks,
Eqs.~\re{Q0-as}, \re{Q0bar-as} and \re{Q1-as}, have to verify the Wronskian
condition \re{Wron}. Taking into account \re{Q1-as}, we obtain after
some algebra the following expression for \re{Wron}%
\footnote{In arriving at this relation we neglected terms suppressed by a small
parameter \re{ksi}.}
\be
\frac{i^Nt_N(u)}{\pi}\bigg[\sin(\pi(s-iu))\cdot \varphi_+(u)\lr{\widebar\varphi_-(u^*)}^*
-\sin(\pi(s+iu)) \cdot\varphi_-(u)\lr{\widebar\varphi_+(u^*)}^*\bigg]=1\,.
\label{W-inter}
\ee
Together with \re{phi-final} and \re{phibar-final}, this leads to the following
relation for the $a-$functions
\be
\sin(\pi(s-iu))\cdot a_+(u)\lr{\widebar a_-(u^*)}^*
-\sin(\pi(s+iu)) \cdot a_-(u)\lr{\widebar a_+(u^*)}^*={\rm const}\,,
\label{W-A}
\ee
which should hold for arbitrary complex $u$.

Another constraint on the $a-$functions comes from Eqs.~\re{Q-finite} and
\re{Q-dec}. It is easy to see that at $(u=i(1-s)+\epsilon,\bar u=-i\bar
s+\epsilon)$ and $(u=-is+\epsilon,\bar u=i(1-\bar s)+\epsilon)$ the blocks
\re{Q0-as}, \re{Q0bar-as} and \re{Q1-as} have poles in $\epsilon$ generated by
the first term involving $\varphi_+-$functions, while the second term
proportional to $\varphi_--$functions is suppressed as $\epsilon^N$. As a
consequence, substituting \re{Q-dec} into \re{Q-finite} we find that
$\varphi_+(u)\bar\varphi_+(\bar u)-\e^{2i\delta}\lr{\varphi_+(\bar u^*)\bar
\varphi_+(u^*)}^*$ has to scale as $\sim \epsilon^N$ as $\epsilon\to 0$ around
the above two points. Applying Eqs.~\re{phi-final} and \re{phibar-final} and
taking into account 
\re{A-period} we get
\be
\e^{2i\delta}=\frac{a_+(\pm is+\varepsilon)\widebar a_+(\pm i\bar s+\varepsilon)}
{\lr{a_+(\pm is+\varepsilon)\widebar a_+(\pm i\bar s+\varepsilon)}^*}+{\cal
O}(\varepsilon^N)\,.
\label{gamma}
\ee
As we will see in a moment (see Eq.~\re{gamma-best}), this relation is exact and
it holds for arbitrary real $\varepsilon$. This implies in particular that
$\delta$ is real in \re{Q-dec}.



So far, we have obtained two different asymptotic expressions for the function
$Q(u,\bar u)$. One of them follows from the WKB expression for the wave function,
$Q(x/\eta,\bar x/\eta)$, Eq.~\re{Q-gen}. Another one follows from \re{Q-dec}
after one replaces the $Q-$blocks by their expressions, Eqs.~\re{Q0-as},
\re{Q0bar-as} and \re{Q1-as}. We remind that the two expressions were obtained
for large values of the charges $q_n$, but in the different regions of
parameters, $x/\eta\gg 1$ and $u$=fixed, respectively. Choosing $u=x/\eta$ and
$\bar u=x^*/\eta$, we notice that Eqs.~\re{Q-gen} and \re{Q-dec} have to coincide
for $x \ll 1$ and $u \gg 1$.

%

To perform the matching, we examine the behaviour of the holomorphic wave
functions $Q_\pm(x/\eta)$ in \re{Q-gen} for $u=x/\eta={\rm fixed}$ as $\eta\to
0$. Using Eqs.~\re{Q-WKB}, \re{dS_0} and \re{dS_1} and choosing $x_0=0$, one
finds after some algebra
\be
Q_\pm(u){\sim} \left[t_N(u) u^{(2s-1)N}\right]^{-1/2}\exp\lr{\pm i\Upsilon(u)}\,,
\label{Q-small-x}
\ee
where the notation was introduced for the function
\be
\Upsilon(u)=u\ln\frac{t_N(u)}{u^N}-\sum_{k=1}^N\lambda_k\ln(\lambda_k-u)
+\sum_{k=1}^N\lambda_k\ln\lambda_k\,,
\label{Upsilon}
\ee
with $\lambda_k$ being the roots of the polynomial $t_N(u)$ defined in \re{roots}
and \re{roots-1}. Notice that $\Upsilon(0)=0$. To obtain $\bar Q_\pm(\bar
x/\eta)$ one has to replace in \re{Q-small-x}, $t_N(u) \to \bar
t_N(u^*)=(t_N(u))^*$, $s\to\bar s=1-s^*$, $\lambda_k \to\bar
\lambda_k=\lambda_k^*$. Then, substituting \re{Q-small-x} into \re{Q-gen} and
making use of \re{A-curve}, one gets
\be
Q(u,u^*)={\rm const}\times|t_N(u)|^{-1}{\exp\lr{-iN\,{\rm Arg\,} (u^{2s-1})}}{}
\cos\lr{2\Re\Upsilon(u)-\Theta}\,,
\label{Q-match}
\ee
where ${\rm Arg\,} (u^{2s-1})\equiv -i\ln(u^{2s-1}{u^*}^{2\bar s-1})/2$. This
expression should match into \re{Q-dec} at large $u$ and $\bar u=u^*$.

Let us find the large$-u$ asymptotics of the $Q-$blocks, Eqs.~\re{Q0-as},
\re{Q0bar-as} and \re{Q1-as}. To stay away from the poles of these blocks on the
complex $u-$plane, we choose $\Re(1-s+iu)>0$. One gets from \re{phi-minus-final}
(up to an inessential overall constant)
\ba
\widehat\varphi_+(u)\, \Gamma^N(1-s+iu)&\sim& \left[t_N(u)
u^{(2s-1)N}\right]^{-1/2}\exp\lr{-i\Upsilon(u)+i\vartheta(u)}\,,
\nonumber
\\
\widehat\varphi_-(u)/{\Gamma^{N}(s+iu)} &\sim& \left[t_N(u)
u^{(2s-1)N}\right]^{-1/2}\exp\lr{i\Upsilon(u)-i\vartheta(u)}\,,
\label{phi-as}
\ea
where $\Upsilon(u)$ is given by \re{Upsilon} and $\vartheta(u)$ is defined as
\be
\vartheta(u)=i\pi uN_-
+\sum_{\Im\lambda_k<0} \lambda_k\ln\lr{i\lambda_k} +\sum_{\Im\lambda_k>0}
\lambda_k\ln\lr{-i\lambda_k}\,,
\label{vartheta}
\ee
with $N_-$ introduced in \re{N_pm}. We observe a striking similarity between
\re{phi-as} and \re{Q-small-x}. Identifying the r.h.s.\ of the first and the
second relations in \re{phi-as} as $Q_\mp(u)\exp(\pm i\vartheta(u))$,
respectively, we get from \re{Q0-as}, \re{Q0bar-as} and \re{Q1-as} the following
expressions for the holomorphic blocks
\ba
Q_0^{\rm (as)}(u)&\sim& a_-(u) Q_+(u)\e^{-i\vartheta(u)}+
\frac{\sin(\pi(s-iu))}{\pi}a_+(u) Q_-(u)\e^{i\vartheta(u)}\,,
\nonumber
\\
Q_1^{\rm (as)}(u)&\sim& \lr{{\widebar a}_-(u^*)}^*
Q_+(u)\e^{-i\vartheta(u)}+\frac{\sin(\pi(s+iu))}{\pi} \lr{\widebar a_+(u^*)}^*
Q_-(u)\e^{i\vartheta(u)}\,,
\label{Q0-asymp}
\ea
and similar expressions for the antiholomorphic blocks
\ba
\widebar Q_0^{\rm (as)}(\bar u)&\sim& \widebar a_-(\bar u)
\widebar Q_-(\bar u)\e^{i(\vartheta(\bar u^*))^*}+
\frac{\sin(\pi(\bar s+i\bar u))}{\pi}\widebar a_+(\bar u) \widebar
Q_+(\bar u)\e^{-i(\vartheta(\bar u^*))^*}\,,
\nonumber
\\
\widebar Q_1^{\rm (as)}(\bar u)&\sim& \lr{{a}_-(\bar u^*)}^*
\widebar Q_-(\bar u)\e^{i(\vartheta(\bar u^*))^*}+\frac{\sin(\pi(\bar s-i\bar u))}{\pi} \lr{a_+(\bar u^*)}^*
\widebar Q_+(\bar u)\e^{-i(\vartheta(\bar u^*))^*}\,.
\label{Q1-asymp}
\ea
The functions $Q_{0,1}^{\rm (as)}(u)$ and $Q_\pm(u)$ approximate the exact
solutions to the holomorphic Baxter equation \re{Bax-eq} in the different
regions, $u\gg 1$ and $u\sim 1$, respectively. Eq.~\re{Q0-asymp} sews these two
sets of functions in the intermediate region of $u$.

Let us substitute Eqs.~\re{Q0-asymp} and \re{Q1-asymp} into \re{Q-dec} and
compare the resulting expression for $Q(u,\bar u)$ with \re{Q-gen} at $u=x/\eta$
and $\bar u=u^*$. One finds that $Q(u,u^*)$ involves four different combinations
of the $Q-$functions, $Q_\pm\widebar Q_\pm$ and $Q_\pm \widebar Q_\mp$, whereas
the r.h.s.\ of \re{Q-gen} contains only the diagonal terms. To cancel the
off-diagonal terms $\sim Q_\pm \widebar Q_\mp$, one has to require that
\be
\e^{2i\delta}=\frac{a_+(u)\widebar a_+(u^*)}{\lr{a_+(u)\widebar a_+(u^*)}^*}=
\frac{a_-(u)\widebar a_-(u^*)}{\lr{a_-(u)\widebar a_-(u^*)}^*}\,.
\label{gamma-best}
\ee
The coefficients $c_\pm$ in front of the diagonal terms $Q_\pm\widebar Q_\pm$ are
given by
\be
c_+=\frac1{\pi}\e^{-2i\Re\vartheta(u)-i\delta}\bigg[{\sin(\pi(\bar
s+iu^*))}a_-(u)\widebar a_+(u^*)
 -\e^{2i\delta}\,{\sin(\pi(\bar s-iu^*))}\lr{a_+(u)\widebar
a_-(u^*)}^*
\bigg]\,,
\label{C-match}
\ee
and $c_-=-c_+^*$. Eq.~\re{C-match} can be further simplified. Replacing $\delta$
by its expressions \re{gamma-best} and making use of the Wronskian relation
\re{W-A}, one gets (up to an overall normalization factor)
\be
c_+=\e^{-2i\Re\vartheta(u)-i\delta}\frac{a_-(u)}{\lr{
a_-(u)}^*}=(-1)^{n_s+n_u}\e^{-2i\Re\vartheta(u)-i\delta}\frac{\widebar a_+(u^*)}
{\lr{\widebar a_+(u^*)}^*}\,.
\label{C-simp}
\ee
Here, in the last relation we applied the identity
$[\sin(\pi(s-iu))]^*=\sin(\pi(1-\bar s+iu^*))=(-1)^{n_s+n_u}\sin(\pi(s-iu))$ with
$n_u=i(u-u^*)$ integer in virtue of \re{x}. Finally, the resulting expression for
$Q(u,u^*$) matches \re{Q-match} provided that the coefficients $c_\pm$,
Eq~\re{C-match}, do not depend on $u$. In addition, these coefficients satisfy
the relation \re{A-curve}, ${c_+}/{c_-}=-c_+/c_+^*=\exp\lr{-2i\Theta}$, which
leads together with \re{C-simp} to
\be
c_\pm= i \e^{\mp i\Theta}
\,,\qquad
\frac{a_-(u)}{\lr{a_-(u)}^*}=(-1)^{n_s+n_u}\frac{\widebar a_+(u^*)}
{\lr{\widebar a_+(u^*)}^*} = i\e^{2i\Re\vartheta(u)+i\delta-i\Theta}
\,.
\label{AA-ratio}
\ee
We recall that these relations hold for $u$ taking the same values as separated
coordinates \re{x}, that is $u=\nu_u-in_u/2$ with $n_u$ integer and $\nu_u$ real.
For a given set of the integrals of motion, the phases $\Re\vartheta(u)$ and
$\Theta$ entering \re{AA-ratio} are uniquely by Eqs.~\re{vartheta} and
\re{Theta}. In contrast, the phase $\delta$ depends on the normalization of the
blocks. Substituting $a_\pm (u) \to \e^{i\delta/2} a_\pm(u)$ and $\bar a_\pm (u)
\to\e^{i\delta/2} \bar a_\pm(u)$ in \re{AA-ratio}, one can put $\delta=0$ in Eqs.~\re{Q-dec}.

Eqs.~\re{AA-ratio} and \re{gamma-best} fix the phases of the $a-$functions but
not their absolute values. Nevertheless, as shown in Appendix~\ref{App:aux}, this
data becomes enough to construct the eigenvalues of the Baxter operator
$Q(u,u^*)$ and to calculate the energy spectrum of the model.


\section{Appendix: Calculation of the energy spectrum}
\label{App:aux}

In this Appendix we obtain the WKB expressions for the energy and quasimomentum,
Eqs.~\re{Energy-1} and \re{quasi-integral}, respectively.

To begin with, we substitute $u=\pm is+\epsilon$ and $\bar u=\pm i \bar
s+\epsilon$ into asymptotic expressions for the $Q-$blocks, Eqs.~\re{Q0-as} and
\re{Q0bar-as}, and examine the limit $\epsilon\to 0$. One finds that the
$\Gamma-$functions in the denominator suppress the contribution of one of the
terms in the r.h.s.\ of \re{Q0-as} and \re{Q0bar-as}. This leads to the following
expression for the quasimomentum \re{Quasimomentum}
\be
\theta_N=i\ln \frac{a_+(is)\widebar a_-(i\bar s)}{\widebar a_+(-i\bar s)a_-(-is)}
+i\ln
\frac{\widehat\varphi_+(is)\lr{\widehat\varphi_-(is-i)}^*}{\widehat\varphi_-(-is)
\lr{\widehat\varphi_+(i-is)}^*}=\theta_N^{(a)}+\theta_N^{(b)}\,,
\ee
where, for convenience, we split the expression into two pieces. One applies
\re{phi-minus-final} and uses the asymptotic behaviour of the $\Gamma-$functions
at large arguments to get
\be
\theta_N^{(b)}=-4\Re\left[\sum_{\Im\lambda_k<0} \lambda_k\ln\lr{i\lambda_k}
+\sum_{\Im\lambda_k>0} \lambda_k\ln\lr{-i\lambda_k}\right]\,.
\label{th-b}
\ee
The calculation of $\theta_N^{(a)}$ goes as follows. One substitutes $u=\pm is$
into \re{W-A} and uses anti-periodicity of the function $\bar a_+(u)$,
Eq.~\re{A-period}, to obtain $a_+(is)\lr{\widebar a_-(i\bar
s)}^*=a_-(-is)\lr{\widebar a_+(-i\bar s)}^*$. This relation allows us to rewrite
$\theta_N^{(a)}$ as
\be
\theta_N^{(a)}=i\ln\frac{\lr{a_-(-is)}^*}{a_-(-is)}
\frac{a_+(is)}{\lr{a_+(is)}^*}=-2\delta+i\ln\frac{\lr{a_-(-is)}^*}{a_-(-is)}
\frac{\lr{\widebar a_+(i\bar s)}^*}{\widebar a_+(i\bar s)}\,,
\ee
where in the second relation we used Eq.~\re{gamma-best}. Then, one takes into
account \re{AA-ratio} and \re{vartheta} to get
\be
\theta_N^{(a)}=-2\Theta+4\Re\left[\sum_{\Im\lambda_k<0} \lambda_k\ln\lr{i\lambda_k}
+\sum_{\Im\lambda_k>0} \lambda_k\ln\lr{-i\lambda_k}\right]\quad ({\rm mod}~
2\pi)\,.
\label{th-a}
\ee
Finally, combining together \re{th-b} and \re{th-a} we obtain
\re{quasi-integral}.

The calculation of the energy \re{Energy} goes along the same lines. One
substitutes \re{Q0-as} and \re{Q0bar-as} into \re{Energy} and expresses the
result in the following form
\be
E_N=\left\{-2\Im\lr{\ln\frac{\widehat\varphi_+(is)}
{[\widehat\varphi_+(i(1-s))]^*}}'+\Delta
E_N\right\}-2\Im\left(\ln\frac{a_+(is)}{\widebar a_+(-i\bar s)}\right)'
=E_N^{(a)}+E_N^{(b)}\,.
\ee
Here, the notation was introduced for an additive constant
\be
\Delta E_N=-2N\Re[\psi(1-2s)+\psi(1-2\bar s)]+\varepsilon_N=-4N\psi(1)\,,
\ee
with $\varepsilon_N$ defined in \re{Energy} and $\Re\psi(1-2\bar
s)=\Re\psi(2s-1)$ in virtue of $\bar s=1-s^*$. To evaluate $E_N^{(b)}$ we apply
\re{gamma} to get
\be
E_N^{(b)}=-\Im\left(\ln\left[\frac{a_+(is)}{\lr{a_+(is)}^*}
\frac{\lr{\widebar a_+(-i\bar s)}^*}{\widebar a_+(-i\bar s)}\right]\right)'
=-\Im\left(\ln\left[\frac{\lr{\widebar a_+(i\bar s)}^*}{\widebar a_+(i\bar s)}
\frac{\lr{\widebar a_+(-i\bar s)}^*}{\widebar a_+(-i\bar s)}\right]\right)'=0\,,
\ee
where the last relation follows from \re{AA-ratio} and \re{vartheta}. Finally,
one rewrites $E_N^{(a)}$ as
\be
E_N^{(a)}=-2\Im\bigg(\ln\left[\widehat\varphi_+(is)\,
\widehat\varphi_+(i(1-s))\right]\bigg)'-4N\psi(1)\,,
\ee
takes into account \re{phi-minus-final} and arrives at \re{Energy-1}.

According to \re{quasi-integral}, the quasimomentum is given by
\be
\theta_N=-\frac{2\pi}{N}\ell\,,\qquad
\ell=\frac{N}{\pi}\Re\int_{P_0^-}^{P_0^+} dx\, S_0'(x) \,,
\label{quas}
\ee
where the ``action'' differential was defined in \re{dS_0}%
\footnote{Here we neglected the last term in the r.h.s.\ of \re{dS_0} since it
does not contribute to \re{quas}.}
\be
\label{action}
dx\,S_0'(x)=\frac{Nt_N(x)-xt'_N(x)}{y(x)}dx\,.
\ee
In Eq.~\re{quas} the integration contour, $\gamma_{P_0^-P_0^+}$, goes on the
Riemann surface \re{curve} from the point $P_0^-$ located above $x=0$ on the
lower sheet to the point $P_0^+$ above $x=0$ on the upper sheet and does not
intersect the cycles $\alpha_k$ and $\beta_k$ $(k=1,..., N-2)$ as shown in
Figure~\ref{Fig-cuts}.

Let us consider the following integral
\be
I_k=\int_{\gamma(\sigma_k)} dx\,S_0'(x)=\int_{\gamma(\sigma_k)} dx
\frac{N  t_N(x)-x  t_N'(x)}{\sqrt{t_N^2(x)-4x^{2N}}}\,.
\label{I-0}
\ee
Here the integration contour $\gamma(\sigma_k)$ goes from the point $P_0^+$ above
$x=0$ on the upper sheet of $\Gamma_N$ to the branching point $\sigma_k$,
encircles it and goes to the point $P_0^-$ on the lower sheet of $\Gamma_N$.
Notice that $\gamma(\sigma_k)$ is different from $\gamma_{P_0^-P_0^+}$, but the
two contours are related to each other through \re{path}. It follows from
\re{quas} that $\ell=N/\pi\Re I_k~ ({\rm mod}~N)$ leading to \re{Re-quas}.


In general, $I_k$ is a complicated function of the integrals of motion
$\Mybf{q}$. The integral in \re{I-0} can be easily evaluated for $\Mybf{q}$
satisfying \re{hierarchy}. In that case, the spectral curve takes the form
\re{Gamma-as} and the branching points are given by \re{br_as}. Replacing the
integration variable in \re{I-0} as $x\to u_N x$ and taking into account the
hierarchy \re{u-moduli} one gets for $j=1,\ldots,N$
\be
I_j=2u_N\int_0^{\e^{{i\pi}(2j-1)/N}}\frac{N\,dx}{\sqrt{1+x^N}}\left[1+
\frac12\lr{\frac{x^N}{1+x^N}-\frac2{N}x\partial_x}p_{N-2}(x)+\mathcal{O}(\epsilon^2)
\right]\,,
\label{J-int}
\ee
where $p_{N-2}(x)=u_2 x^{N-2}+...+u_{N-1}x=\mathcal{O}(\epsilon)$ and we
neglected terms quadratic in $p_{N-2}(x)$. Straightforward calculation leads to
\be
I_j=2u_N\left[\e^{{i\pi}(2j-1)/N}{\rm B}\lr{\frac12,\frac1{N}}+
\frac1{N}\sum_{n=2}^{N-1}\e^{{i\pi}(2j-1)n/N} u_{N+1-n}\,{\rm B}\lr{\frac12,\frac{n}{N}}
+\mathcal{O}(\epsilon^2)\right]\,,
\label{J-as}
\ee
with ${\rm B}(x,y)=\Gamma(x)\Gamma(y)/\Gamma(x+y)$ being the Euler function. This
expression has the following properties
\be
I_{j+N}=I_{j}\,,\qquad\sum_{j=1}^N I_{j} =0+\mathcal{O}(\epsilon^{3/2})\,.
\label{J-prop}
\ee
The r.h.s.\ of \re{J-as} takes the form of a discrete Fourier transformation from
the coordinate ($j$) to the momentum ($n$) representation. This allows one to get
the moduli $u_N$ and $u_n$ as inverse Fourier transformation of $I_j$. Then,
taking into account that $\Re I_k=\pi(n_k+\ell/N)$, Eq.~\re{Re-quas}, one arrives
at \re{Fourier}.

\section{Appendix: Calculation of the quasimomentum}

Let us demonstrate that for the integrals of motions $\Mybf{q}$ satisfying the
quantization conditions \re{quan-cond}, the parameter $\ell$ takes strictly
integer values in \re{quas}. To begin with, we define on the Riemann surface
\re{curve} the set of normalized differentials of the first kind, $\omega_k$, and
the third kind (dipole differentials), $\Omega_\infty$,
\be
\label{exp-diff}
\omega_k =\sum_{j=1}^{N-2} U_{kj}\frac{dx x^{j-1}}{y}\,,\qquad
\Omega_\infty=2q_2^{1/2}\frac{dx x^{N-2}}{y}+\sum_{j=1}^{N-2} U_{j}\frac{dx
x^{j-1}}{y}\,,
\ee
with the expansion coefficients $U_{kj}$ and $U_{j}$ fixed by the normalization
conditions~\cite{D}
\be
\label{diff}
\oint_{\alpha_j} \omega_k=2\pi \delta_{jk}\,,\qquad
\oint_{\alpha_j} \Omega_\infty=0\,.
\ee
The differential $\Omega_\infty$ has a pair of poles located at the points
$(P_\infty^-,P_\infty^+)$ above $x=\infty$ on the upper and lower sheets of
$\Gamma_N$ and the residue at these poles $\res_{P_\infty^\pm} \Omega_\infty =\pm
1$. The action differential \re{action} can be decomposed over the set of the
differentials \re{exp-diff} as
\be
\label{dS-dec}
dx\,S_0'(x)= q_2^{1/2}\cdot\Omega_\infty +\sum_{k=1}^{N-2}{a_k}
\cdot\omega_k\,.
\ee
Here, the coefficient in front of $\Omega_\infty$ is fixed by the asymptotic
behaviour of $dxS_0'(x)$, Eq.~\re{action}, at infinity $S_0'(x\to
P_\infty^\pm)\sim \pm q_2^{1/2}/x$. In the second term, the notation was
introduced for the $\alpha-$periods of the action differential
\be
\label{a-period}
a_k=\frac1{2\pi}\oint_{\alpha_k} dx\,S_0'(x)\,,\qquad \Re a_k=\frac12
\ell_{2k-1}\,,
\ee
whose values are fixed by the quantization conditions \re{quan-cond}.

Substituting \re{dS-dec} into \re{quas}
and applying the well-known identities between the contour integrals on the
Riemann surface \cite{D}
\be
\label{iden}
\int_{P_0^-}^{P_0^+}\Omega_\infty=\int_{P_\infty^-}^{P_\infty^+}\Omega_0\,,\qquad
\int_{P_0^-}^{P_0^+}\omega_k=-i\oint_{\beta_k}\Omega_0\,,
\ee
one can express $\ell$
in terms of the integrals of the dipole differential $\Omega_0$
\be
\label{quas-2}
\ell=\frac{N}{\pi}\Re\left[q_2^{1/2}\int_{P_\infty^-}^{P_\infty^+}\Omega_0
-i\sum_{k=1}^{N-2} {a_k} \oint_{\beta_k}\Omega_0\right]\,.
\ee
The differential $\Omega_0$ has a pair of poles located at the points $P_0^\pm$
on $\Gamma_N$ above $x=0$ and is normalized as $\oint_{\alpha_j} \Omega_0=0$.
This differential plays a special role in our analysis as it can be expressed in
terms of the quasimomentum
\be
\label{Omega0}
\Omega_0=-\frac1{N} dx\,p'(x)\,,\qquad \e^{p(x)}+\e^{-p(x)}=\frac{t_N(x)}{x^N}\,.
\ee
Notice that $p(x)$ is a multi-valued function on the Riemann surface \re{curve}
such that
\be
\label{special}
\e^{p(P_\infty^\pm)}=1\,,\qquad \e^{p(\sigma_k)}=\pm 1\,,
\ee
with $\sigma_k$ being the edges of the cuts, $t_N^2(\sigma_k)=4\sigma_k^{2N}$. As
a consequence,
\be
\label{integrals}
\int_{P_\infty^-}^{P_\infty^+}\Omega_0=2\pi i \frac{m}{N}\,,\qquad
\oint_{\beta_k}\Omega_0=2\pi i \frac{m_k}{N}
\ee
%
%
with $m$ and $m_k$ integer. Finally, we find from \re{quas-2}
\be
\label{final}
\ell= {2N} \Re\left[i \frac{m}{N}q_2^{1/2} +\sum_{k=1}^{N-2} a_k
\frac{m_k}{N} \right]
=nm +2\sum_{k=1}^{N-2}m_k \Re {a_k} = nm+\sum_{k=1}^{N-2} m_k\ell_{2k-1}\,.
\ee
Here, in the second relation we took into account that $ q_2^{1/2}=i(h-1/2)+{\cal
O}\lr{(h-1/2)^{-1}} ={in}/2-\nu\,, $ with $h=(1+n)/2+i\nu$. Notice that the
quasimomentum depends only on the $\alpha-$periods of the action differential.

\end{document}